\documentclass[twocolumn,dvipsnames]{aastex61}
\usepackage{amsfonts,amsmath,amssymb,bm}
\usepackage{booktabs}
\usepackage{fontenc}
\usepackage{graphicx} 
\usepackage{amssymb}      
\usepackage{mathrsfs}      
\usepackage{array}       
\usepackage{longtable}     
\usepackage{natbib}       
\usepackage{multirow,makecell}
\usepackage{lineno}

\newcounter{rownumbers}

           \newcommand{\sGV}{DeVore\_GV}
           \newcommand{\sSA}{DeVore\_SA}
           \newcommand{\sKM}{DeVore\_KM}
           \newcommand{\sJT}{Coulomb\_JT}
           \newcommand{\sSY}{Coulomb\_SY}

\newcommand{\CBSP}{CB$_{\rm SP}$}
\newcommand{\CBFF}{CB$_{\rm FF}$}
\newcommand{\CFITsc}{CFIT$_{\mathrm sc}$}
\newcommand{\FYL}{FI$_{\rm YL}$}
\newcommand{\FEP}{FI$_{\rm EP}$}

\newcommand{\Nabla}{\vec{\nabla}}
\newcommand{\Times}{\vec{\times}}
\newcommand{\rmd}{{\rm d}}

\newcommand{\dv}{\, \rmd \mathrm{V}}
\newcommand{\ds}{\, \rmd \mathrm{S}}
\newcommand{\surf}{{\partial \mathcal{V}}}
\newcommand{\vol}{\mathcal{V}}

\newcommand{\ints}{\int_{\surf}}
\newcommand{\intv}{\int_{\vol}}
\newcommand{\curlA}{\Nabla \times \vA}
\newcommand{\curlAp}{\Nabla \times \vAp}
\newcommand{\divA}{\Nabla \cdot \vA}
\newcommand{\divAp}{\Nabla \cdot \vAp}

\newcommand{\curlB}{\Nabla \times \vB}
\newcommand{\divB}{\Nabla \cdot \vB}
\newcommand{\divBp}{\Nabla \cdot \vBp}

\newcommand{\vA}{\vec{A}}
\newcommand{\vAp}{\vA_{\rm p}}
\newcommand{\vB}{\vec{B}}

\newcommand{\Bz}{B_{\rm z}}
\newcommand{\Blos}{B_{\rm los}}
\newcommand{\vBp}{\vB_{\rm p}}

\newcommand{\vJ}{\vec{J}}

\newcommand{\Bt}{\vec{B}_{\mathrm t}}
\newcommand{\Bn}{B_{\mathrm n}}
\newcommand{\vt}{\vec{v}_t}
\newcommand{\vn}{v_n}
\newcommand{\vp}{v_{\perp}}
\newcommand{\vu}{\vec{u}}
\newcommand{\vx}{\vec{x}}

\newcommand{\E}{E}          
\newcommand{\ECB}{E_{\rm CB}}
\newcommand{\Ep}{E_{\rm p}}      
\newcommand{\EpCB}{E_{\rm p,CB}}
\newcommand{\Ej}{E_{\rm J}}      
\newcommand{\Eps}{E_{\rm p,s}}    
\newcommand{\Ejs}{E_{\rm J,s}}    
\newcommand{\Emix}{E_{\rm mix}}    
\newcommand{\EdivBJ}{E_{\rm J,ns}}  
\newcommand{\EdivBp}{E_{\rm p,ns}}  
\newcommand{\Epns}{E_{\rm p,ns}}   
\newcommand{\Ejns}{E_{\rm J,ns}}   
\newcommand{\Etot}{E}   
\newcommand{\Efree}{E_{\rm F}}   
\newcommand{\EfreeCB}{E_{\rm F,CB}} 
\newcommand{\Ediv}{E_{\rm div}}    
\newcommand{\Edivprime}{E_{\rm div}/E}    
\newcommand{\Emixprime}{|E_{\rm mix}|/\Ejs} 

\newcommand{\aff}{\alpha_{\rm ff}}
\newcommand{\Hdef}{\mathscr{H}_\vol}
\newcommand{\Hdefm}{\langle\mathscr{H}_\vol\rangle}
\newcommand{\Hdefmabs}{\langle|\mathscr{H}_\vol|\rangle}
\newcommand{\HdefJ}{\mathscr{H}_{\vol,J}}
\newcommand{\HdefJm}{\langle\mathscr{H}_{\vol,J}\rangle}
\newcommand{\HdefJP}{\mathscr{H}_{\vol,JP}}
\newcommand{\HdefJPm}{\langle\mathscr{H}_{\vol,JP}\rangle}
\newcommand{\HdefCB}{\mathscr{H}_{\rm CB}}
\newcommand{\HdefCBlh}{{\mathscr{H}}_{\rm CB,LH}}
\newcommand{\HdefCBrh}{{\mathscr{H}}_{\rm CB,RH}}
\newcommand{\THdef}{\tilde{\mathscr{H}}_\vol}

\newcommand{\hjprime}{|\HdefJ|/|\Hdef|}
\newcommand{\hjprimem}{\langle|\HdefJ|/|\Hdef|\rangle}
\newcommand{\Hacc}{H_{\vol,\rm acc}}
\newcommand{\Haccs}{H_{\vol,\rm acc}^\prime}

\newcommand{\avfi}[1]{\langle\,|f_i#1|\,\rangle}

\newcommand{\epsN}{\epsilon_{\rm N}}
\newcommand{\epsM}{\epsilon_{\rm M}}

\newcommand{\tj}{\theta_J}

\newcommand{\strtable}{\renewcommand{\arraystretch}{1.2}} 

\newcommand{\gauss}{{\rm G}}

\newcommand{\mx}{{\rm Mx}}
\newcommand{\mxmx}{{\rm Mx$^2$}}
\newcommand{\mxmxs}{{\rm Mx$^2$\,s$^{-1}$}}
\newcommand{\erg}{{\rm erg}}
\DeclareMathAlphabet{\mathbfit}{OML}{cmm}{b}{it}
\renewcommand{\vec}{\mathbfit}
\newcommand{\asecs}{\mbox{\ensuremath{^{\prime\prime}}}}
\accepted{August 19, 2021}
\submitjournal{APJ}
\received{June 25, 2021}

\shorttitle{Magnetic Helicity: Method Comparison on Solar Observations}
\shortauthors{Thalmann et al.}

\begin{document}

\title{Magnetic helicity estimations in models and observations of the solar
magnetic field. Part IV: Application to solar observations}

\author{J.\ K.\ Thalmann} 
\affil{University of Graz, Institute of Physics/IGAM, Graz, Austria}
\author{M.\ K.\ Georgoulis} 
\affil{Research Center for Astronomy and Applied Mathematics of the Academy of Athens, Athens, Greece}
\author{Y.\ Liu} 
\affil{W.\ W.\ Hansen Experimental Physics Laboratory, Stanford, CA, USA}
\author{E.\ Pariat} 
\affil{Laboratoire de Physique des Plasmas (LPP), CNRS, Sorbonne Universit\'e, \'Ecole polytechnique, Institut Polytechnique de Paris, Palaiseau, France}
\affil{LESIA, Observatoire de Paris, Universit\'e PSL, CNRS, Sorbonne Universit\'e, Universit\'e de Paris, Meudon, France}
\author{G.\ Valori} 
\affil{Max-Planck-Institut f\"ur Sonnensystemforschung, G\"ottingen, Germany}
\author{S.\ Anfinogentov} 
\affil{Institute of Solar-Terrestrial Physics, Irkutsk, Russia}
\author{F.\ Chen} 
\author{Y.\ Guo} 
\affil{School of Astronomy and Space Science, Nanjing University, Nanjing, China}
\author{K.\ Moraitis} 
\affil{Physics Department, University of Ioannina, Ioannina, Greece}
\author{S.\ Yang}
\affil{Key Laboratory of Solar Activity, National Astronomical Observatories, Chinese Academy of Sciences, Beijing, China}
\collaboration{\footnotesize(The ISSI Team on Magnetic helicity)}
\nocollaboration
\author{A.\ Mastrano} 
\affil{Sydney Institute for Astronomy, School of Physics, University of Sydney, NSW, Australia}

\begin{abstract}
In this ISSI-supported series of studies on magnetic helicity in the Sun, we systematically implement different magnetic helicity calculation methods on high-quality solar magnetogram observations. We apply finite-volume, discrete flux tube (in particular, connectivity-based) and flux-integration methods to data from Hinode's Solar Optical Telescope. The target is NOAA active region 10930 during a ~1.5 day interval in December 2006 that included a major eruptive flare (SOL2006-12-13T02:14X3.4). Finite-volume and connectivity-based methods yield instantaneous budgets of the coronal magnetic helicity, while the flux-integration methods allow an estimate of the accumulated helicity injected through the photosphere. The objectives of our work are twofold: A cross-validation of methods, as well as an interpretation of the complex events leading to the eruption. To the first objective, we find (i) strong agreement among the finite-volume methods, (ii) a moderate agreement between the connectivity-based and finite-volume methods, (iii) an excellent agreement between the flux-integration methods, and (iv) an overall agreement between finite-volume and flux-integration based estimates regarding the predominant sign and magnitude of the helicity. To the second objective, we are confident that the photospheric helicity flux significantly contributed to the coronal helicity budget, and that a right-handed structure erupted from a predominantly left-handed corona during the X-class flare. Overall, we find that the use of different methods to estimate the (accumulated) coronal helicity may be necessary in order to draw a complete picture of an active-region corona, given the careful handling of identified data (preparation) issues, which otherwise would mislead the event analysis and interpretation.
\end{abstract}


\section{Introduction}
\label{s:intro} 

\subsection{Relative helicity and its estimation}

Magnetic helicity is a signed scalar quantity that numbers the structural complexity of a magnetic field. For a given volume, it is written in the form
\begin{linenomath*}
\begin{equation}
\Hdef \equiv \intv \left(\vA\cdot\vB \right) \dv,
\label{eq:h}
\end{equation}
\end{linenomath*}
where $\vB=\curlA$ and $\vA$ corresponds to the magnetic vector potential. The integral form of \href{eq:h}{Eq.~(\ref{eq:h})} represents a generalization of the definition of magnetic helicity based on the winding number that quantifies the linkage of a discrete number of magnetic field lines/flux tubes \citep{1969JFM....35..117M}. Magnetic helicity has the property of being exactly conserved in ideal MHD, and quasi-conserved even in resistive magneto-hydrodynamics (MHD) in the case of a high magnetic Reynolds number \citep[e.g.,][]{1984GApFD..30...79B}. As a result, it has been suggested to represent a fundamental physical driver of coronal mass ejections (CMEs), in order to balance the otherwise impossible-to-dissipate total solar helicity production \citep[e.g.,][]{1994PhPl....1.1684L, 1994GeoRL..21..241R}.

To make the concept of magnetic helicity applicable to arbitrary magnetic field distributions, a divergence-free ($\divB=0$) magnetic field must be bounded by a magnetically closed volume, namely, $\vB\cdot\vec{\hat{n}}|_\surf=0$, where $\surf$ is the boundary of the volume $\vol$. The latter condition, however, is hard to achieve in natural systems such as the solar corona. For this reason, and also because magnetic fields in the solar atmosphere thread the photospheric boundary, the concept of ``relative'' magnetic helicity has been introduced by \cite{1984JFM...147..133B} and \cite{1985CPPCF...9...111F} in the form
\begin{linenomath*}
\begin{equation}
\Hdef \equiv \intv \left(\vA+\vAp\right) \cdot \left(\vB-\vBp\right) \dv,
\label{eq:hr}
\end{equation}
\end{linenomath*}
where $\vB$ and $\vBp$ are generated by vector potentials $\vA$ and $\vAp$, respectively, and $\vBp$ is a reference magnetic field. A current-free (i.e., potential) magnetic field is commonly used as reference. Such a field is defined by
\begin{linenomath*}
\begin{equation}
\vBp = \nabla \varphi,
\label{eq:bp}
\end{equation}
\end{linenomath*}
where $\varphi$ is a scalar potential that satisfies $(\vec{\hat{n}}\cdot\nabla\varphi)|=(\vec{\hat{n}}\cdot\vB)|$ on $\surf$ \citep[for an alternative choice of the reference field see, e.g.,][]{2020ApJ...894..151Y}. Together with $\divB=\divBp=0$, $\Hdef$ in \href{eq:hr}{Eq.~(\ref{eq:hr})} is gauge-invariant, i.e., it represents a physically meaningful quantity that can be used to characterize the magnetic system within $\vol$ \citep[e.g.,][]{2012SoPh..278..347V}. It furthermore represents a well-conserved quantity in ideal and resistive MHD \citep{2015A&A...580A.128P,2013SoPh..283..369Y,2018ApJ...865...52L,2018A&A...613A..27Y}. For brevity, we hereafter use the term magnetic helicity to refer to the relative magnetic helicity.

The application of \href{eq:hr}{Eq.~(\ref{eq:hr})} to the solar corona is hampered by several difficulties. Central among them is our inability to measure the coronal magnetic field reliably on a routine basis \citep[for a review see, e.g.,][]{2009SSRv..144..413C}. Therefore, for a given volume of interest, the coronal magnetic field is typically approximated by a nonlinear force-free (NLFF) field within a finite volume (FV), which requires the routinely measured surface vector magnetic field as the lower boundary condition \citep[for reviews see, e.g.,][]{2012LRSP....9....5W,2017SSRv..210..249W}. Using the 3D model magnetic field as input, the FV helicity based on \href{eq:hr}{Eq.~(\ref{eq:hr})} can be readily evaluated once $\vec{A}$ and $\vec{A}_{\rm p}$ are known. Different methods have been developed to compute the vector potentials in Cartesian geometry \citep[e.g.,][]{2011SoPh..272..243T,2012SoPh..278..347V,2013SoPh..283..369Y,2014SoPh..289.4453M}.

The impact of the specific NLFF magnetic field model for the analysis of coronal magnetic energy and relative helicity budget is yet to be fully assessed. In a first comparative analysis of the dependence of FV helicity estimates on the spatial resolution of the underlying NLFF models, \cite{2015ApJ...811..107D} found that obtaining consistent estimates is a challenging, yet achievable task. More precisely, given a certain FV helicity method, the obtained values of the coronal helicity differed by a factor of five at most (see their Table~2 and Fig.~8).

Once successfully computed, $\Hdef$ may be further decomposed into two separately gauge-invariant expressions, namely $\Hdef=\HdefJ + \HdefJP$ \citep{1999PPCF...41B.167B}, where
\begin{linenomath*}
\begin{eqnarray}
\HdefJ &\equiv& \intv \left(\vA-\vAp\right) \cdot \left(\vB-\vBp\right) \dv,
\label{eq:hj}\\
\HdefJP &\equiv& 2 \intv \vAp \cdot \left(\vB-\vBp\right) \dv.
\label{eq:hjp}
\end{eqnarray}
\end{linenomath*}
Here, $\HdefJ$ loosely represents the helicity of the current-carrying part of the considered magnetic field $\vB_{\rm J}=\vB-\vBp$ (called ``current-carrying'' helicity, hereafter), and $\HdefJP$ represents the part of helicity associated with the field threading the boundaries of $\vol$ (called ``volume-threading'' 
helicity, hereafter). Despite being independently gauge invariant, $\HdefJ$ and $\HdefJP$ are not conserved in ideal MHD (in contrast to $\Hdef$ defined in \href{eq:hr}{Eq.~(\ref{eq:hr})}) due to the existence of a gauge-invariant transfer term that enables the exchange between the two terms \citep[][]{2018ApJ...865...52L}. 
Recent attention has been drawn to $\HdefJ$ in \href{eq:hj}{Eq.~(\ref{eq:hj})} as this term provides additional information compared to $\Hdef$. In particular, the so-called ``(non-potential) helicity ratio'', $\hjprime$, appeared as a promising metric of the eruptive potential of the studied magnetic structure. This was noted not only in numerical simulations \citep[e.g.,][]{2017A&A...601A.125P,2018ApJ...863...41Z,2018ApJ...865...52L}, but also in observationally-based studies \citep[][]{2018ApJ...855L..16J,2019A&A...628A..50M,2019ApJ...887...64T,2019A&A...628A.114P}.

Alternatively to the FV methods mentioned above, some helicity-calculation approaches rely on the representation of the magnetic field as a collection of discrete, finite-sized flux tubes within a FV. Such methods will be hereafter referred to as discrete flux-tubes (DT) methods, and include the twist-number (TN) method \citep[][]{2017ApJ...840...40G} and the connectivity-based (CB) method \citep{2012ApJ...759....1G}. Among the discrete methods, the TN method requires full knowledge of the magnetic field in $\vol$, while the CB method relies on the lower (photospheric) boundary only, modeling an optimal coronal connectivity based on it.

Besides requiring the full three-dimensional magnetic field, the TN method \citep{2010ApJ...725L..38G,2013ApJ...779..157G} requires a magnetic flux rope to be present in the volume, in order to relate its twist with the helicity. A flux rope is a magnetic structure that has attracted strong interest in recent decades and consists briefly of a significantly twisted magnetic field winding around a relatively untwisted, or less twisted, axis \citep[for reviews and definitions, see][]{titov_demoulin99,gibson_etal06,priest2014}
The CB method, on the other hand, models the coronal field as a single \citep[linear;][]{2007ApJ...671.1034G}, or a collection of \citep[nonlinear;][]{2012ApJ...759....1G} flux tube(s). For details on the CB method, see \href{ss:dt_methods}{Sect.~\ref{ss:dt_methods}}.

In \citet{2016SSRv..201..147V}, existing FV, DT and TN methods have been reviewed, bench-marked and assessed in terms of performance. In that comprehensive work, a variety of numerical magnetic configurations were studied, considered to be relevant for solar magnetic helicity studies. The considered test configurations differed in their topological complexity, the magnitude and spatial distribution of electric currents in the model volumes (large-scale smoothly distributed vs.\ localized direct currents), as well as their stability properties (in the form of snapshots of time-dependent non-force-free MHD simulations of flux emergence). We summarize their findings in \href{ss:scope}{Sect.~\ref{ss:scope}}, in relation to the scope of the present study.

The helicity in a given volume, $\vol$, may also be interpreted as resulting from a net helicity flux through the bounding surface $\surf$, e.g., using an helicity flux equation such as \citep[see][for other formulations]{2015A&A...580A.128P}:
\begin{linenomath*}
\begin{equation}
\frac{\mathrm{d}\Hdef}{\mathrm{d}t}= 2 \ints \left[ \left(\vAp \cdot \Bt \right) \vn - \left(\vAp \cdot \vt \right) \Bn \right] \ds.
\label{eq:dhdt}
\end{equation}
\end{linenomath*}
This applies for a specified set of conditions on the vector potentials, in the time interval, say, $T=\int _0^T {\mathrm d}t$ \citep[e.g.,][]{1984GApFD..30...79B,1999PPCF...41B.167B} in the absence of helicity dissipation \citep[][]{1984JFM...147..133B}. Here, $\Bt$ and $\Bn$ denote the tangential and normal magnetic field components, respectively, while $\vt$ and $\vn$ are the respective tangential and normal components of the velocity $\vp$ perpendicular to the magnetic field $\vB$. Notice also that the reference field $\vBp$ and $\vB$ have identical normal components on $\surf$. Once the magnetic and velocity fields on $\surf$ are known, \href{eq:dhdt}{Eq.~(\ref{eq:dhdt})} can be readily implemented. Its first term is sometimes called "emergence" or "advection" term, as it is associated with $\vn$. The second term of \href{eq:dhdt}{Eq.~(\ref{eq:dhdt})} is sometimes called ``shear'' term, as it is associated with $\vt$. Note however that these terms are gauge dependent and their intensities can change when different gauges are used (cf.\ examples in \citet{2015A&A...580A.128P,2018ApJ...865...52L}), which makes their physical interpretation as separate quantities disputable.

Upon application to the solar atmosphere, one has to assume that the bounding surface $\surf$ in \href{eq:dhdt}{Eq.~(\ref{eq:dhdt})} represents the solar photosphere permeated by the helicity flux that determines the helicity content in the coronal volume above. For a finite (Cartesian) volume this implies that the helicity flux through the lateral and top boundaries of $\vol$ is assumed to be negligible. To evaluate \href{eq:dhdt}{Eq.~(\ref{eq:dhdt})}, the velocity field vector has to be deduced from time series of photospheric magnetic field observations (i.e., magnetograms), obtained on a routine basis. The principle of several velocity inversion methods have been reviewed by \citet{2007ApJ...670.1434W}. Deriving the velocity field is a nontrivial task, as it involves temporal derivatives of the measured surface magnetic field components, radial and/or tangential. Hence, the quality of the resulting velocity fields relies, on top of the velocity estimation methods, on the magnetogram quality and cadence \citep[cf.][]{2007ApJ...670.1434W}.

\cite{2003SoPh..215..203D} showed that it is possible to simplify the expression for the helicity flux across the photospheric boundary, by evaluating
\begin{linenomath*}
\begin{equation}
\frac{\mathrm{d}\Hdef}{\mathrm{d}t} = -2 \ints \left(\vAp \cdot \vu \right) \Bn \ds,
\label{eq:dhdt_simple}
\end{equation}
\end{linenomath*}
where $\vu=\vt-(\vn/\Bn)\Bt$ is the flux transport velocity, which corresponds to the apparent transverse velocity of the footpoints of elementary flux tubes. The flux transport velocity can be theoretically derived using velocity inversion methods from time series of magnetograms \citep[eg.][]{2007ApJ...670.1434W,2008ApJ...683.1134S}. However, it remains unclear to which extent any velocity inversion method when applied to observational data is able to measure the true flux transport velocity, hence the real photospheric helicity flux \citep[eg.][]{2008ApJ...683.1134S,2009AdSpR..43.1013D,2012ApJ...761..105L}.

In brief, the so-called helicity-flux integration (FI) methods follow the time evolution of the photospheric magnetic field to determine the variation of accumulated coronal helicity with respect to an unknown initial state (see \href{ss:fi_methods}{Sect.~\ref{ss:fi_methods}} for details). Some of the existing FI methods \citep{2005A&A...439.1191P,2012ApJ...761..105L} are reviewed in a forthcoming work \citep{2021SSRv..vol..pageP}.

\subsection{Context and scope of this study}
\label{ss:scope}

Along with \citet{2016SSRv..201..147V}, \citet{2017ApJ...840...40G}, \citet{2021SSRv..vol..pageP}, the present work is part of a series of works carried out by the ISSI team on "Magnetic Helicity estimations in models and observations of the solar magnetic field"\footnote{\url{http://www.issibern.ch/teams/magnetichelicity/index.html}} that aims to compare and benchmark different methods to measure relative magnetic helicity. The seminal paper of the series, \citet{2016SSRv..201..147V}, provides a review of different helicity measurement methods, mainly focused on testing different FV methods, based on physically meaningful test magnetic fields (semi-analytical test setups and snapshots of 3D MHD numerical experiments). They demonstrated that all but one of the seven tested FV methods gave reliable and consistent results, mutually agreeing to within 3\%.

The high level of agreement between the FV methods was reached when the magnetic field was sufficiently solenoidal, i.e., if $\divB=0$ was sufficiently well respected. Using a dedicated test, \citet{2016SSRv..201..147V} proposed a respective threshold above which helicity estimates lack reliability. \citet{2019ApJ...880L...6T} explicitly showed the unpredictable effect of insufficient solenoidality of NLFF models onto subsequent FV helicity computation, but also demonstrated the ability of two different FV methods to provide consistent helicity estimates given that the NLFF models are sufficiently solenoidal \citep[see also][]{2019ApJ...887...64T}. The first major objective of the present study is thus to complete these earlier studies by performing the first systematic comparison of multiple methods on observation-based data, in order to verify that consistent results can be obtained.

\citet{2016SSRv..201..147V} also compared the helicity estimates from application of the CB and TN methods to that of the FV methods, using synthetic data sets. They found that for a flux emergence simulation mimicking a stable (non-eruptive) corona, the CB method provided a helicity estimate agreeing to within $\approx10\%$ with that of the FV methods. Yet for a different simulation of an eruptive (CME-productive) corona, the CB method was significantly underestimating the FV helicity, being different by a factor of 2 -- 8. Moreover, it appeared that the CB method works better for sufficiently complex 2D magnetic configurations. Since observational data can be better approximated by a collection of flux tubes (as thought for in the CB method), we may therefore expect the CB method to provide helicity estimates more consistent with that of FV methods. Regarding the TN method, \citet{2016SSRv..201..147V} and \citet{2017ApJ...840...40G} showed that it is capable of measuring the helicity carried by the current-carrying part of the magnetic field, thus of $\HdefJ$ in \href{eq:hj}{Eq.~(\ref{eq:hj})}. Thus, another aspect of the present study is to complete the analysis of \citet{2016SSRv..201..147V}, this time using observation-based data for the comparison of FV helicity estimates with those of the CB and TN methods.

\citet{2021SSRv..vol..pageP} tests different FI methods on data from 3D MHD numerical experiments of solar-like phenomena (both, of eruptive and non-eruptive type) and found that only when applied properly and carefully, consistent results are obtained (with an agreement to within $\approx$1\%). A comparison to the corresponding FV-based results showed that the FI methods provide helicity estimates of a simulated (CME-productive) corona, agreeing to within $\approx$20\% during the non-eruptive phase. In contrast, timely around the simulated solar-like eruption, the FI methods and FV methods expectantly deliver strongly different results. Thus, another aspect of the present study includes thus to purse a corresponding analysis using observational data. An important difference in respect to similar earlier works is that we also perform a thorough analysis of the effects of the particular data (calibration) on the retrieved helicity fluxes, allowing us to question earlier findings.

Finally, the second major objective of the present work is to provide a better understanding and a more complete physical insight of the evolution of the magnetic helicity (and thus the underlying magnetic field) of the studied coronal magnetic system. This is achieved by combining the helicity estimates of the different approaches noted above, each being based on a different hypothesis and subject to a different scientific purpose. In particular, we study NOAA active region (AR) 10930 in the time interval 2006 December~11 -- 13, that includes an eruptive X3.4 flare (SOL2006-12-13T02:14X3.4) and a full-halo CME \citep[e.g.,][]{2011ApJ...740...68F,2016ApJ...824...93F}. To perform the analysis, we rely on high-quality photospheric vector magnetograms (\href{s:data}{Sect.~\ref{s:data}}) and the state of the art methods for NLFF coronal magnetic field modeling (\href{ss:data_fv}{Sect.~\ref{ss:data_fv}}) and helicity computation (\href{s:h_methods}{Sect.~\ref{s:h_methods}}).

\section{Data}
\label{s:data}

\begin{figure*}[htb]
\centerline{\centering\includegraphics[width=0.64\textwidth]{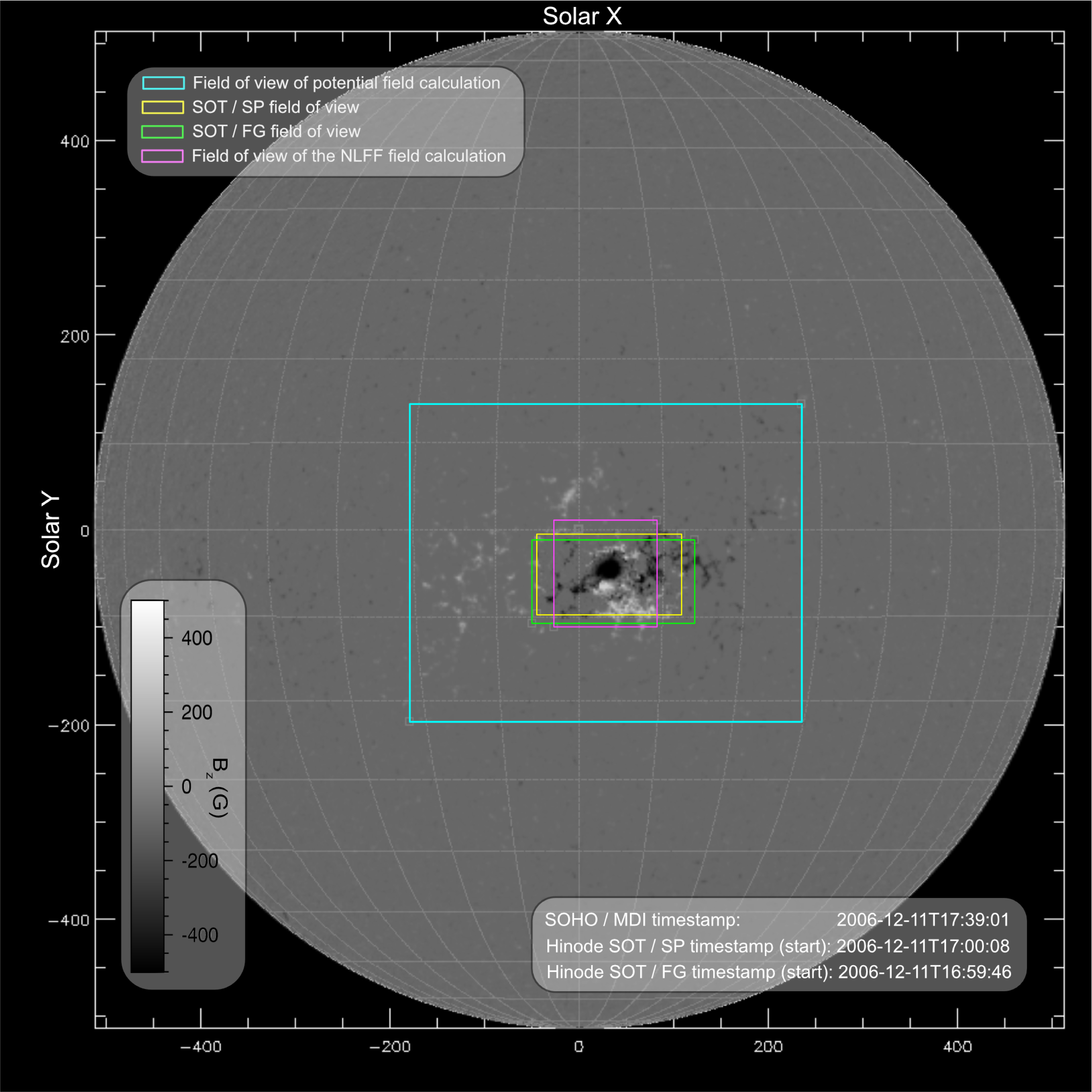}}
\caption{Full-disk synthetic solar magnetogram on 2006 December 11 at around 17:00 UT, including NOAA AR~10930. The global photospheric field is provided by SOHO/MDI. Shown as rectangles are the different FOVs of studied magnetograms at this time: the one on which the photospheric potential field vector was calculated (cyan), the one on which the NLFF field extrapolation was applied (magenta) and the two Hinode SOT FOVs, namely the SOT-NFI (green) and SOT-SP (yellow). The latter FOV includes the embedded SOT-SP image. The synthetic magnetogram is saturated at $\pm 500$~\gauss. \\}
\label{fig:fov}
\end{figure*}

In the following, the sources and processing particular data used for NLFF modeling and/or helicity computations are discussed. A summarizing \href{tab:data}{Table~\ref{tab:data}} can be found in \href{s:data_append}{Appendix~\ref{s:data_append}}.

\subsection{Vector magnetogram data for NLFF modeling}
\label{ss:data_nlff}

The Solar Optical Telescope \citep[SOT;][]{2008SoPh..249..167T} Spectro-Polarimeter \citep[SP;][]{2013SoPh..283..579L} on board the \emph{Hinode} spacecraft \citep{2007SoPh..243....3K} operates in a fixed wavelength band centered on the Zeeman-sensitive Fe\,{\sc i} lines at 6302~\AA. SOT-SP obtained vector magnetogram sequences of NOAA AR 10930 for over a week, with a near-continuous coverage.

We used Level-2 SOT SP data, available at \url{https://csac.hao.ucar.edu/sp_data.php}. For the FV method and the NLFF extrapolations we used three magnetograms obtained at 17:00~UT on December~11, 20:30~UT on December~12, and 04:30~UT on December~13, 2006, respectively. At these times, the AR was located around W04$^\circ$/S05$^\circ$ (see \href{fig:fov}{Fig.~\ref{fig:fov}}), W18$^\circ$/S05$^\circ$, and W23$^\circ$S05$^\circ$, respectively. Given its relative proximity to the disk center, the magnetograms did not exhibit substantial projection effects.

Notice that for the NLFF magnetic field reconstruction on December~12 and 13, we use as input the same magnetic fields as in \cite{2008ApJ...675.1637S}. The main steps taken in preparation of the input vector magnetic field data are summarized in the following \citep[see also Sect.~2 of][for more details]{2008ApJ...675.1637S}. In a first step in \cite{2008ApJ...675.1637S}, Level-1.5 SP vector magnetic field data \citep[][and references therein]{2007MmSAI..78..148L} were subjected to a minimum-energy (ME) azimuth disambiguation \citep[]{1994SoPh..155..235M,2006SoPh..237..267M}. The disambiguated SP vector data were embedded into a much larger, lower-resolution \emph{SOHO}/MDI line-of-sight (LOS) magnetogram in order to incorporate larger-scale flux information, beyond the SP field-of-view (FOV). The data were binned by a factor of two, to a plate-scale of $0.63\asecs$.

For the December~11 NLFF modeling in this study, we applied a procedure designed to provide input data as consistent as possible with the two magnetograms of December 12 and 13. In particular, we prepared a homogeneous Level-2 SP data set acquired by the SP scan modes between 17:00:08~UT and 18:03:20~UT on December~11. A nearly simultaneous full-disk \emph{SOHO}/MDI \citep{1995SoPh..162..129S} LOS magnetogram (\href{fig:fov}{Fig.~\ref{fig:fov}}) was interpolated by a factor of three, to an effective pixel size of $0.66\asecs$. The SP data were then binned to the pixel size of the embedding MDI data by means of synthetic Stokes images that were then inverted to provide the binned magnetograms. These magnetograms were disambiguated using the non-potential magnetic field calculation (NPFC) method of \citet{2005ApJ...629L..69G}, as refined in \citet{2006SoPh..237..267M}. Notice that the NPFC azimuth disambiguation method used for the December 11 magnetogram is different than the ME method used for the December 12 and 13 magnetograms. The reason for this choice is twofold: first, the comparison of the FV and CB methods on magnetograms disambiguated via two different methods (Section \ref{sss:cbff_results}) and, second, the correspondence with the SOT-SP magnetograms to which the CB helicity calculation method was applied (Sections \href{ss:data_cb16}{Sections~\ref{ss:data_cb16}} and \href{sss:cbsp_results}{\ref{sss:cbsp_results}}), that were also disambiguated using the NPFC method. The NPFC-disambiguated and binned SP magnetogram of December 11 (yellow outline in \href{fig:fov}{Fig.~\ref{fig:fov}}) was then embedded into the binned MDI magnetogram (cyan outline). Finally, a sub-field was selected for the NLFF analysis, covering a photospheric area nearly identical (in terms of the area physically covered) to that of the already available December~12 and 13 magnetic field maps (magenta outline).

\subsection{Vector magnetic field data for FV computations}
\label{ss:data_fv}

In order to be able to compute the relative helicities from \href{eq:hr}{Equations~(\ref{eq:hr})}, \href{eq:hj}{(\ref{eq:hj})}, and \href{eq:hjp}{(\ref{eq:hjp})}, we apply the individual FV methods described in \href{ss:fv_methods}{Sect.~\ref{ss:fv_methods}} to $\vB$ and $\vBp$ obtained via NLFF modeling, as explained in the following.

The NLFF field in and above NOAA~10930 was reconstructed using the procedure described in \cite{2009ApJ...700L..88W}, representing an optimization of the Grad-Rubin method (``CFIT'') introduced by \cite{2007SoPh..245..251W} (see \href{s:nlff_append}{Appendix~\ref{s:nlff_append}} for details). This method is favored in the present work because of a number of advantageous properties. These include the method's strict convergence to a single, self-consistent force-free solution (therefore referred to as ``\CFITsc'', hereafter), achieved by successive averaging of the individual contributing maps of the force-free parameter alpha (one for the positive-polarity and one for the negative-polarity subdomains). A further favorable property is to achieve an accurate solution to the force-free problem when applied to solar data, including a high degree of solenoidality (see \href{s:quality_append}{Appendix~\ref{s:quality_append}} for details).

In the present work, we used the SOT-SP vector magnetic field data described in \href{ss:data_nlff}{Sect.~\ref{ss:data_nlff}} (for its footprint on the solar disk see the magenta outline in \href{fig:fov}{Fig.~\ref{fig:fov}}) as input to the \CFITsc\ method, where electric currents in weak field regions ($<5\%$ of the maximum field strength) were censored out and corresponding uncertainties assumed as $\propto1/|B_z|^2$. For completeness, we note that the effect of censoring on the original SOT-SP data is largest for the December~11 data set ($\approx 3\%$ of the total unsigned magnetic flux) and is negligible for the December~12 and 13 data sets.

The computational volume for the \CFITsc\ models of December~12 and 13 covers $320^2\times256$~pixel, with a plate scale of $0.63\asecs$. The model volume for December~11, given the slightly different spatial resolution of the input magnetic field data of $0.66\asecs$, was accordingly set as $305^2\times244$~pixel, covering the same approximate coronal volume. We note here that all \CFITsc\ models satisfy generally-used metrics regarding their force-freeness and level of solenoidality (divergence-freeness), justifying their subsequent use for helicity computations (see \href{tab:metrics}{Table~\ref{tab:metrics}}).

Besides the requirements on the solenoidal quality of the magnetic fields, $\vB$ and $\vBp$, discussed in \href{ss:scope}{Sect.~\ref{ss:scope}}, the vector potentials $\vA$ and $\vAp$ required in the computation of $\Hdef$ from \href{eq:hr}{Eq.~(\ref{eq:hr})} must reproduce the respective input magnetic field as accurately as possible. Therefore, we apply the metrics introduced in \cite{2006SoPh..235..161S} to the pairs ($\vB$, $\curlA$) and ($\vBp$, $\curlAp$) for each of the considered FV methods, and list them for the interested reader in \href{tab:ei_hi}{Table~\ref{tab:ei_hi}}. 

\subsection{Vector magnetogram data for CB estimates}
\label{ss:data_cb16}

In the present case, the CB method is applied to two data sets: first, to the \CFITsc\ lower boundary data described in \href{ss:data_fv}{Sect.~\ref{ss:data_fv}}. This will provide the \CBFF\ helicity estimation that will be directly compared to the FV estimates. Second, to a series of SOT-SP magnetograms, described in this section. This second use of the CB method provides the \CBSP\ estimation of helicity that is also compared to the FV measurements.

On top of the three SOT-SP magnetograms selected for MDI insertion, another 13 Level-2 vector magnetograms acquired between 11~December $\sim$03:10~UT and 13~December $\sim$16:21~UT were selected and processed for the application of the \CBSP\ method. These magnetograms, with the exception of three, are included in the SOT-SP database with a spatial sampling of $\sim$0.31~arcsec per pixel. The other three magnetograms are included at full resolution of $\sim$0.16~arcsec in the database and were binned by a factor of two for homogeneity with the rest of the data series. The observation times of all 16 magnetograms, along with the results of the analysis, are included in \href{tab:cbsp_detailed}{Table~\ref{tab:cbsp_detailed}}.

All these SOT-SP magnetograms were disambiguated using the NPFC method. As explained in \citet{2005ApJ...629L..69G}, disambiguation is performed on the local (i.e., de-projected) magnetic field components on the image (i.e., observation) plane. The disambiguated magnetograms were then co-aligned to determine a common FOV. Although the CB method is applied to each magnetogram independently, a common FOV helps to mitigate against inconsistencies in the pseudo-times series of the results that are due to flux patches included in some, but not all, magnetogram maps. The results shown in this study have been obtained from these local magnetic field components on which disambiguation was applied.

\subsection{Magnetogram and flux transport velocity data for FI computations}
\label{ss:data_fi}

FI methods primarily estimate the flux of magnetic helicity through the photospheric boundary (\href{eq:dhdt_simple}{Eq.~\ref{eq:dhdt_simple}}) which requires the knowledge of the distribution of both, the normal component of the magnetic field, $B_n$, and the flux transport velocity $\vu$. The latter can be obtained from time series of magnetograms thanks to velocity inversion methods \citep[for a review of these methods see][]{2007ApJ...670.1434W}.

Several velocity inversion methods are solely using $B_n$ as input to estimate $\vu$, such as, e.g., the Differential Affine Velocity Estimator \citep[DAVE;][]{2006ApJ...646.1358S}, but also vector magnetograms can be used \citep[e.g., via the Differential Affine Velocity Estimator for Vector Magnetograms (DAVE4VM);][]{2008ApJ...683.1134S}. While \href{eq:dhdt_simple}{Eq.~(\ref{eq:dhdt_simple})} does not explicitly require vector magnetograms as an input, the derivation of $\vu$ can nonetheless benefit from the knowledge of the three components of $\vB$. The supplementary information provided by the additional field components enables a better inversion of the induction equation and therefore a more accurate estimate of $\vu$ \citep[see][]{2008ApJ...683.1134S}. For observation-based applications, \cite{2012ApJ...761..105L} have shown that helicity flux calculations based on the flux transport velocity inferred either by DAVE or by DAVE4VM, were giving very consistent results (within about $20\%$).

However, since the helicity estimates from FI methods result from a time integration, data with a high temporal cadence is needed to accurately picture the corresponding helicity flux evolution. Unfortunately, vector magnetic field data is not acquired with the same time cadence as that of the LOS component only. Thus, there is always a trade-off in using vector magnetic field to deduce $\vu$: while the computation of the helicity flux is likely improved on the one hand, the monitoring of the accumulation of helicity is greatly reduced on the other hand. Hence the usage of LOS data is frequently privileged. It is important to note, however, that irrespective of the particular data source used, velocity inversion methods are far from being able to provide an exact estimation of $\vu$, given the inherent constraints of the magnetic field measurements, and thus are a considerable source of uncertainty in the retrieved helicity fluxes \citep[cf.][]{2007AdSpR..39.1674D,2007ApJ...670.1434W,2008ApJ...683.1134S,2009AdSpR..43.1013D}.

{\catcode`\&=11
\gdef\2018AandA...613A..27Y{\cite{2018A&A...613A..27Y}}}
{\catcode`\&=11
\gdef\12005AandA...439.1191P{\cite{2005A&A...439.1191P}}}

\begin{table*}[t]
\caption{Summary of helicity computation methods implemented in this work, their requirements and deliverables as described in \href{s:h_methods}{Sect.~\ref{s:h_methods}}, their acronym, their main bibliographic reference, as well as their pertinent sections.}
\centering
\footnotesize
\strtable
\begin{tabular}{|c c c c |}
\hline
Requirements and &  Acronym &  Original publication & Appearance in this work\\
main deliverables & & & (Method summary, Results) \\
\hline
\multicolumn{4}{|c|}{\bf Finite volume (FV) helicity}\\
\multirowcell{2}{%
	\makecell[{{>{\parindent -7em}p{5cm}}}]{%
		-- Requires $\vB$ in $\vol$ at one time instant (from\\
		 \hspace{4pt} NLFF  modeling in this work; see \href{ss:data_fv}{Sect.~\ref{ss:data_fv}}). \\ 
		-- Provides instantaneous estimate of $\Hdef$, from \\
		\hspace{4pt} evaluating \href{eq:hr}{Eq.~(\ref{eq:hr})}, and of the individual con-\\
		\hspace{4pt} tributions to it (\href{eq:hj}{Eqs.~(\ref{eq:hj})} and \href{eq:hjp}{(\ref{eq:hjp})}).
	}
} 
& \sJT & \cite{2011SoPh..272..243T} &  \href{ss:fv_methods}{Sect.~\ref{ss:fv_methods}}, \href{ss:hv_results}{Sect.~\ref{ss:hv_results}}\\
& \sSY & \2018AandA...613A..27Y  & -- `` --\\
& \sGV &\cite{2012SoPh..278..347V} & -- `` --\\
& DeVore\_KM &\cite{2014SoPh..289.4453M}  &  -- `` --\\
& DeVore\_SA & \cite{2016SSRv..201..147V}  &  -- `` --\\
\hline
\multicolumn{4}{|c|}{\bf Discrete flux tube (DT) helicity}\\
\multirowcell{1}{%
	\makecell[{{>{\parindent -7em}p{5cm}}}]{%
		-- Requires $\vB$ on $\surf$ at one time instant  (from \\ 
		 \hspace{4pt} NLFF  models or SOT-SP data; see \href{ss:data_cb16}{Sect.~\ref{ss:data_cb16}}). \\ 
		-- Models the coronal connectivity as a collection \\
		\hspace{4pt} of force-free flux tubes.\\
		-- Provides instantaneous estimate of $\Hdef$, based\\
		\hspace{4pt} on a minimal connection length principle.\\
	}
} 
& & & \\
& \CBFF & \cite{2012ApJ...759....1G}  &  \href{ss:dt_methods}{Sect.~\ref{ss:dt_methods}}, \href{sss:cbff_results}{Sect.~\ref{sss:cbff_results}}\\
& \CBSP & -- `` --  &    \href{ss:dt_methods}{Sect.~\ref{ss:dt_methods}}, \href{sss:cbsp_results}{Sect.~\ref{sss:cbsp_results}}\\
& & & \\
& & & \\
\hline
\multicolumn{4}{|c|}{\bf Flux-integration (FI) helicity}\\
\multirowcell{1}{%
	\makecell[{{>{\parindent -7em}p{5cm}}}]{%
		-- Requires time evolution of $\vB$ on $\surf$. \\ 
		-- Requires time evolution of $\vu$ on $\surf$. \\ 
		-- Provides instantaneous estimate of $\frac{\mathrm{d}\Hdef}{\mathrm{d}t}$.\\
		-- Allows to evaluate the accumulation of \\
		\hspace{4pt}  helicity, by time integration of \href{eq:dhdt_simple}{Eq.~(\ref{eq:dhdt_simple})}. \\
	}
} 
& & & \\
&  \FEP & \12005AandA...439.1191P  & \href{ss:fi_methods}{Sect.~\ref{ss:fi_methods}}, \href{ss:h_flux}{Sect.~\ref{ss:h_flux}}\\
&  \FYL & \cite{2012ApJ...761..105L} &  -- `` -- \\
& & & \\
\lasthline
\label{tab:methods}
\end{tabular}
\end{table*}

Since the cadence of the available SOT-SP vector data (see first column in \href{tab:cbsp_detailed}{Table~\ref{tab:cbsp_detailed}}) is too sparse for the purpose of FI computations, we use LOS magnetic field data ($\Blos$) from the SOT Narrowband Filter Imager \citep[NFI;][]{2008PFR.....2S1009I}. The NFI provides polarimetric imaging at high spatial resolution for Fe~lines having a range of sensitivity to the Zeeman effect, centered at 5250~\AA. A series of 1151 LOS magnetograms, spanning the time range 11~December 12:09:20~UT to 13~December 12:59:41~UT, with a time cadence of $\sim$2~minutes and covering the approximate same FOV as the vector magnetograms used by the CB method, were used for FI computations (green outline in \href{fig:fov}{Fig.~\ref{fig:fov}}). The NFI data were calibrated following \cite{2007PASJ...59S.619C}. They suggest, first, the usage of a linear relation between the circular polarization and $\Blos$ in order to calibrate the data outside of umbral areas (their Eq.~7). Second, in order to model the reversal of the polarization signal over field strength in umbral regions (where the ratio of the intensity to the average intensity of the quiet Sun is $<0.35$), a first-order polynomial is suggested (their Eq.~8). This step, however, introduces discontinuities at the boundary of the umbral areas, resulting in unrealistic velocity estimates from DAVE. Therefore, our preferred choice is to use DAVE velocities inferred from the field calibrated in step one above, and to use $\Blos$ after additional application of step two above. For a detailed comparative analysis of the effect of data calibration see \href{ss:calib_append}{Appenndix~\ref{ss:calib_append}}.

Careful inspection of the calibrated NFI data, however, exhibit artifacts spatially related with the saturated areas inside of the main sunspot's umbra during the flare, and related artificially large DAVE velocities. For all FI computations, we therefore exclude the time range 13~December 02:14~UT -- 02:57~UT (the nominal GOES flare duration) from analysis (for more details see \href{ss:disclosure_append}{Appendix~\ref{ss:disclosure_append}}).

\section{Helicity computation methods}
\label{s:h_methods}

In this section, we introduce the individual helicity computation methods used in this work. A guiding list of these methods with related synoptic information can in found in \href{tab:methods}{Table~\ref{tab:methods}}.

\subsection{Finite volume (FV) helicity}
\label{ss:fv_methods}

The FV methods implemented in this study have been reviewed, bench-marked, and their performance assessed in \cite{2016SSRv..201..147V}. The methods can be grouped into Coulomb ($\divA=0$, $\divAp=0$) and DeVore ($A_z=0$, $A_{{\rm p},z}=0$) methods, according to the gauge in which the vector potentials are computed \citep[see Sect.\ 2.1 and 2.2, respectively, in][]{2016SSRv..201..147V}. The Coulomb methods include that of \cite{2011SoPh..272..243T} (``\sJT'') and \cite{2018A&A...613A..27Y} (``\sSY''). The DeVore methods include that of \cite{2012SoPh..278..347V} (``\sGV'') and \cite{2014SoPh..289.4453M} (``DeVore\_KM''), and in addition the ``DeVore\_SA'' implementation described in detail in Sect.~2.2.3 of \cite{2016SSRv..201..147V}.

\subsection{Discrete flux tube (DT) helicity}
\label{ss:dt_methods}

In comparison to the FV methods, the TN method \citep{2010ApJ...725L..38G,2017ApJ...840...40G} performs a parametric fitting of a flux rope (assumed to exist within $\vol$), thus delivers an estimate of the helicity $H_{\rm twist}$ associated to the twist of that structure. \citet[][]{2016SSRv..201..147V} and \citet{2017ApJ...840...40G} have shown that, for cases with a high degree of twist in a present flux rope, the TN method delivers an accurate estimate of the twist, and thus of $\HdefJ$ in \href{eq:hj}{Eq.~(\ref{eq:hj})}. 

The CB method \citep{2012ApJ...759....1G} relies on a multi-polar partitioning of the photospheric flux distribution to approximate the unknown magnetic connectivity in the coronal volume in the form of a collection of slender magnetic flux tubes. Each flux tube is force-free, with a constant force-free parameter related to an average total electric current and flux of a given modeled connection (i.e., flux tube). The ensemble of flux tubes is inferred by prioritizing connections along photospheric magnetic polarity inversion lines by means of simulated annealing. The result is a 'skeletal' NLFF method that delivers a very fast, relatively to the FV methods, lower-limit estimate of the instantaneous magnetic energy and helicity budgets for a (any) local-scale, perfectly flux-balanced, 'connected' flux distribution. Given the lack of a detailed coronal linkage, the CB method ignores energy and helicity terms due to the winding of different flux tubes, assuming simple 'arch-like' tubes instead \citep[see also][for a complete theoretical framework]{demoulin_etal06}.

The CB method was designed with practical applications in mind, ready to be applied to any given photospheric vector magnetogram limited enough to allow Cartesian geometry (cylindrical- or spherical-geometry generalizations are also feasible, but not yet implemented). It only uses the full photospheric magnetic field vector as input. In this sense, it does not fully share the purpose of FV methods which, ideally, are capable of recovering the true value of the relative helicity in a volume at the price of requiring the full three-dimensional field vector in this volume. Being discrete, the CB method also allows the independent calculation of mutual and self free energy and helicity terms, along with the left-handed (LH; $\HdefCBlh$) and right-handed (RH; $\HdefCBrh$) contributions to the total helicity.

The CB method is applied to the \CFITsc\ lower boundary data described in \href{ss:data_fv}{Sect.~\ref{ss:data_fv}}, with results referred to as ``\CBFF'', hereafter. It is also applied to the 16 SOT-SP vector magnetograms described in Section \ref{ss:data_cb16}, with results referred to as ``\CBSP'' hereafter.

\subsection{Flux-integration (FI) helicity}
\label{ss:fi_methods}

The FI methods compute the time integration of the photospheric flux of relative helicity (\href{eq:dhdt_simple}{Eq.~\ref{eq:dhdt_simple}}). Thus, these methods evaluate the accumulation of helicity due to photospheric contributions, instead of directly evaluating the instantaneous helicity content in the coronal domain. The FI methods used in this study, include that of \citeauthor{2005A&A...439.1191P} (\citeyear{2005A&A...439.1191P}; ``\FEP'', hereafter) and \citeauthor{2012ApJ...761..105L} (\citeyear{2012ApJ...761..105L}; ``\FYL'' hereafter) and were applied to the high-cadence LOS magnetic field data described in \href{ss:data_fi}{Sect.~\ref{ss:data_fi}} (for its footprint on the solar disk see green outline in \href{fig:fov}{Fig.~\ref{fig:fov}}). The \FYL\ method directly evaluates \href{eq:dhdt_simple}{Eq.~(\ref{eq:dhdt_simple}}), using $B_n$ given by the observations, $\vu$ being derived using DAVE (cf.\ \href{ss:data_fi}{Sect.~\ref{ss:data_fi}}), and the vector potential $\vAp$ computed using a FFT method with the Coulomb gauge.

The \FEP\ method evaluates a different version of \href{eq:dhdt_simple}{Eq.~(\ref{eq:dhdt_simple}}), assuming that $\vAp$ satisfies the Coulomb gauge. \citet{2005A&A...439.1191P} demonstrated that \href{eq:dhdt_simple}{Eq.~(\ref{eq:dhdt_simple})} is equivalent to:
\begin{linenomath*}
\begin{equation}
  \frac{\mathrm{d}\Hdef}{\mathrm{d}t} = \ints \ints \frac{\Bn \Bn'((\vu-\vu')\Times(\vx-\vx'))_n}{2\pi(\vx-\vx')^2} \ds \ds'.
  \label{eq:dhdt_gtheta}
\end{equation}
\end{linenomath*}
The \FEP\ method directly computes the helicity flux from $B_n$ and $\vu$. Assuming that the magnetic field distribution can be represented by a collection of elementary magnetic elements, the \FEP\ method estimates the magnetic flux weighted relative rotation of all pairs of elementary magnetic elements \citep{2005A&A...439.1191P}. The \FEP\ method requires the numerical computation of a double integral, making it more resource demanding than the \FYL\ method when applied to magnetic field data of high spatial resolution.

These \FEP\ and \FYL\ methods have been tested and benchmarked in \citet{2021SSRv..vol..pageP} on synthetic datasets and shown to deliver an agreement of deduced helicity fluxes with high precision, deviating by a few percent only. However, both methods evaluate the helicity flux only where magnetic data are available, i.e., on the limited physical area covered by the studied magnetograms. As a consequence, the helicity flux stemming from outside of the studied area, i.e., that which penetrates the corresponding coronal volume through the lateral and top boundaries, is therefore necessarily neglected. Based on synthetic modeling of an isolated emerging solar-like active region, \citet{2021SSRv..vol..pageP} showed that the FI methods are able to recover the coronal helicity content fairly well when the dynamics of the system was non-eruptive: the relative difference between the FI and FV method where of $\sim0.5\%$, $\sim 10\%$ and $\sim20\%$ for the three simulations studied in \citet{2021SSRv..vol..pageP}.

Typically, when applied to observed magnetic field data, FI computations are only carried out for data points within the magnetograms where the magnitude of $\Bz$ is larger than a certain threshold, in order to reduce the computation time. This is justified since weak magnetic field is known to contribute only little to the overall photospheric helicity flux. In order to be able to address the effect of this thresholding, we carry out helicity flux computations using two different limits: 20~\gauss\ (representing a typical noise level of the SOT-NFI measurements) and 100~\gauss. In addition, the \FYL\ method, excludes pixels where $\Bz$ is lower than a certain threshold (1$\sigma$), in order to avoid data that is not reliable. 

We apply the relatively fast \FYL\ method to the full time series of SOT-NFI data, once using each of the aforementioned thresholds. Because of the higher computational needs of the \FEP\ method, we run this method only once using a threshold of 100~\gauss, allowing us to compare the relative performance of the two methods, given identical model parameter settings.

\section{Results}
\label{s:results}

\subsection{Coronal magnetic field}
\label{ss:results_cfit}

\begin{figure}
\centerline{\centering\includegraphics[width=0.88\columnwidth]{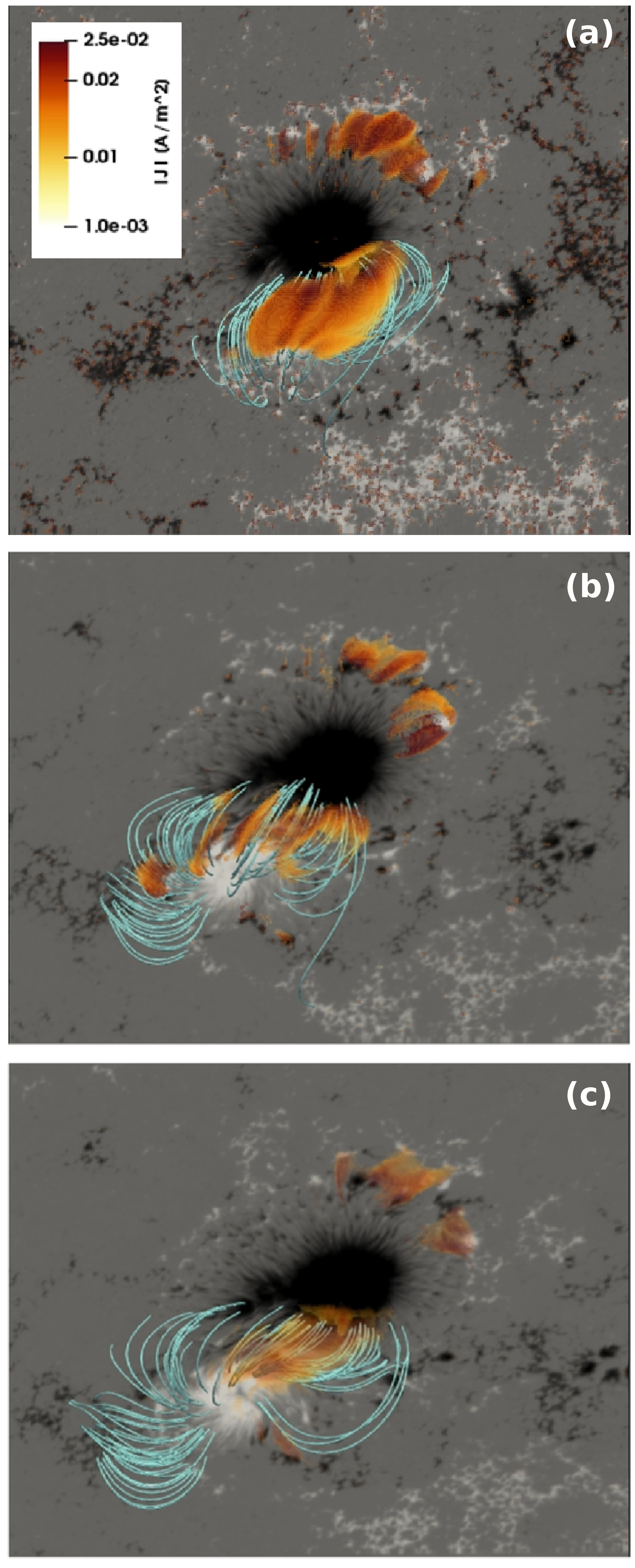}}
\caption{Sample field lines computed from the \CFITsc\ magnetic field models of NOAA AR~10930 on (a) 2006 December~11 at 17:00~UT, (b) December~12 at 20:30~UT, and (c) December~13 at 04:30~UT. Field lines are computed randomly from locations near the polarity inversion line of the active region. The red-shaded volume rendering depicts the places of strongest absolute current density in the range 0.01--0.025~A\,m$^{-2}$. The gray scale background shows the photospheric vertical field $B_z$, saturated at $\pm$2~kG. Black/white color represents negative/positive magnetic polarity, respectively.}
\label{fig:sample_fls}
\end{figure}

The morphology of the pre-flare corona on December~12 at 20:30~UT can be described as a low-lying sheared arcade, connecting the two main sunspots. We also find strong electric currents in the arch filament system, flowing between the main sunspots on December~12 (see \href{fig:sample_fls}{Fig.~\ref{fig:sample_fls}(b)}). These currents in the sheared arcade are found to be weaker in the post-flare snapshot on December~13 (compare \href{fig:sample_fls}{Fig.~\ref{fig:sample_fls}(c)}), supporting that a part of pre-existing current density was dissipated. A system of stronger and higher elevating electric currents is found for the December~11 snapshot, although apparently less sheared compared to the two following time instances (\href{fig:sample_fls}{Fig.~\ref{fig:sample_fls}(a)}). 

Based on our \CFITsc\ magnetic field models, we find the highest total unsigned magnetic flux for the December~11 snapshot ($\phi=5.69\times10^{22}$~\mx; see \href{tab:nlff_modeling}{Table~\ref{tab:nlff_modeling}}), larger by a factor of $\sim$1.5 compared to the December~12 and 13 snapshots. Observed strong shearing motions and flux cancellation near the polarity inversion line, spatially separating the two main sunspots \citep[see, e.g., movie associated to Fig.~2 of][]{2008ApJ...675.1637S} are partly responsible for the decrease of unsigned flux between December~11 and 12, and is captured by all methods presented here, as well as in the vector magnetogram data to which \CFITsc\ modeling is applied to. We emphasize here again that we use the vector magnetic field data originally used in \cite{2008ApJ...675.1637S} for the December~12 and 13 snapshots, and that we prepared the input data for the December~11 snapshot as consistently as possible. Nevertheless, some of the particular steps taken in our data preparation do deviate from those in \cite{2008ApJ...675.1637S}  (for details see \href{ss:data_nlff}{Sect.~\ref{ss:data_nlff}}), thus may partially be responsible for the obtained differences in unsigned magnetic flux between the December~11 and 12 snapshots.

\begin{table}
\caption{Values of the unsigned magnetic flux ($\phi$), total ($\E$), potential ($\Ep$) and free ($\Efree=\E-\Ep$) magnetic energies, deduced from the \CFITsc\ models. Mean coronal relative helicity, $\Hdefm$, deduced from the results of the individual FV methods (cf.\ \href{tab:fv_methods_detailed}{Table~\ref{tab:fv_methods_detailed}}). Units of magnetic fluxes, magnetic energies and relative helicities are $10^{22}$\,\mx, $10^{33}$\,\erg\ and $10^{43}$\,\mxmx , respectively.}
\centering
\footnotesize
\strtable
\begin{tabular}{@{~}c c@{\quad} c@{\quad} c@{\quad} c@{\quad} c@{\quad} c@{\quad}}
\hline
Date \& Time & $\phi$ & $\E$ & $\Ep$ & $\Efree$ & $\Hdefm$ \\
\hline
11 Dec 17:00~UT & 5.69 & 2.97 & 2.61 & 0.36 & -2.79$\pm$0.20 \\
12 Dec 20:30~UT & 3.87 & 1.89 & 1.85 & 0.04 & -0.61$\pm$0.04\\ 
13 Dec 04:30~UT & 3.87 & 2.06 & 1.94 & 0.12 & -1.32$\pm$0.08\\
\hline
\label{tab:nlff_modeling}
\end{tabular}
\end{table}

Based on our \CFITsc\ magnetic field models, we find highest magnetic energies on 11~December 17:00~UT, followed by the post-flare NLFF field on 13~December 04:30~UT and the pre-flare configuration on 12~December 20:30~UT. The free magnetic energy ($\Efree$) at those times comprises about 14\%, 6\% and 2\% of $\Ep$, respectively. Correspondingly, we find a higher value of $\Efree$ for the post-flare corona on 13~December 04:30~UT, being about $1\times10^{32}$~\erg\ larger than $\Efree$ on 12~December 20:30~UT.

The consistency of the \CFITsc\ extrapolations as solutions to the NLFF equations is commonly quantified by the degree of force- and divergence-freeness (solenoidality). Corresponding standard measures and their discussion are given in \href{s:quality_append}{Appendix~\ref{s:quality_append}}. In the context of magnetic helicity computations, \cite{2016SSRv..201..147V} showed that the degree of divergence-freeness of the tested field, $\vB$, is one of the key factors that critically determines the spread in the deduced helicity across different FV methods. Comparing the relevant metrics of our \CFITsc\ models (\href{tab:metrics}{Table~\ref{tab:metrics}}) with those of the test cases reported in Table~7 of \cite{2016SSRv..201..147V}, we are confident about the sufficient solenoidal quality of our \CFITsc\ models. For instance, considering the \CFITsc\ model on December~12, the values listed in \href{tab:metrics}{Table~\ref{tab:metrics}} show the sum of all non-solenoidal contributions to amount to $\approx$0.26\% of the total energy. In comparison, \cite{2016SSRv..201..147V} reports for a test case with a similar level of solenoidality, a spread of less than 1\% in the corresponding helicity values from the application of different FV methods (see their Section~7), i.e., was found smaller than differences due to the numerical accuracy of individual FV implementations (as large as 4\%). 

\subsection{Finite-volume helicity}
\label{ss:hv_results}

\subsubsection{Extensive helicities}
\label{sss:hv_extensive}

\begin{figure}
\centerline{\includegraphics[width=0.9\columnwidth]{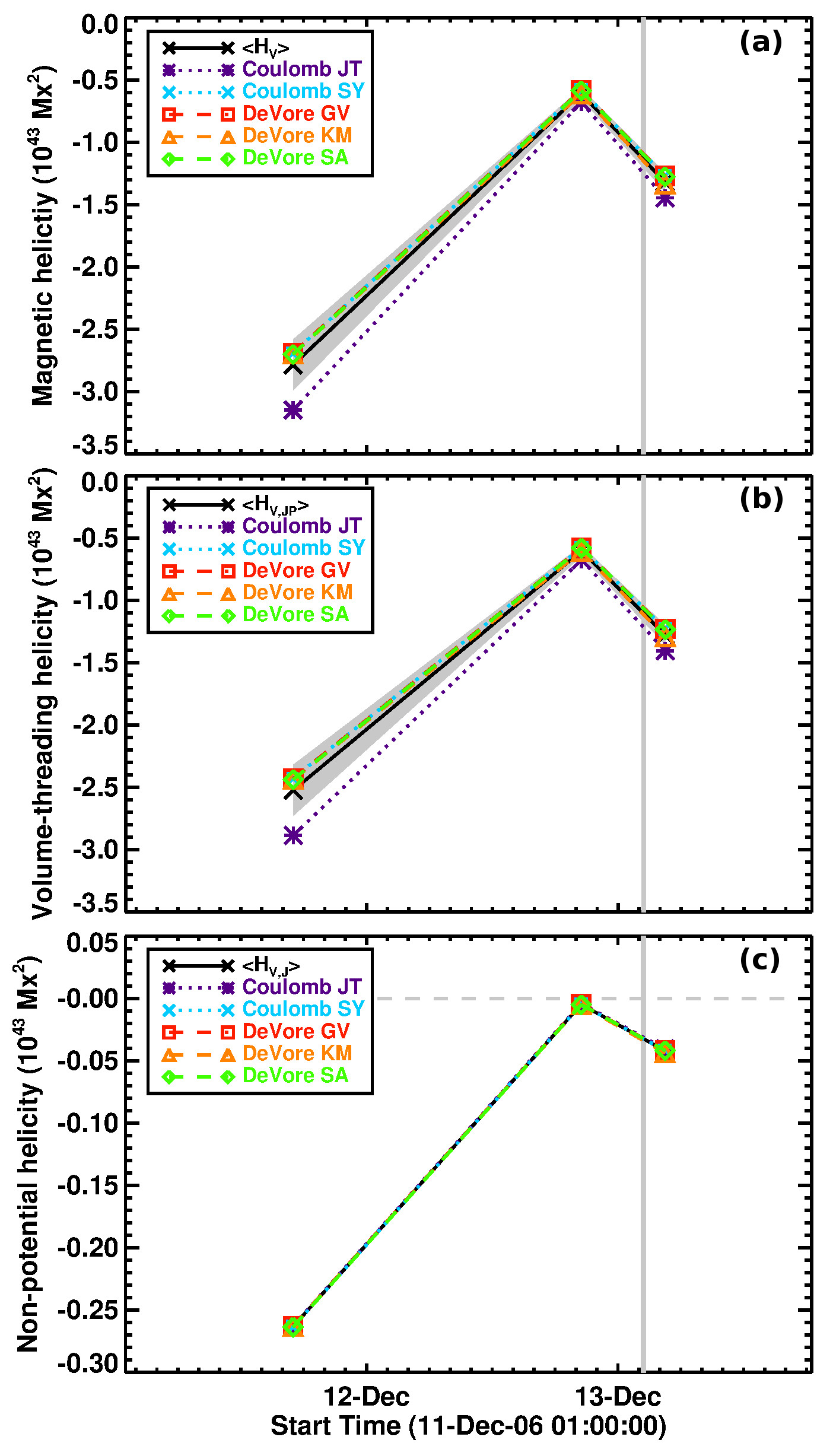}}
\caption{FV helicities for NOAA AR~10930 during 2006 December 11--13. (a) Relative helicitiy ($\Hdef$) computed from the different FV methods. (b) Volume-threading helicity ($\HdefJP$). (c) Current-carrying helicity ($\HdefJ$). Mean values are represented by black crosses (black solid lines). Corresponding standard deviations are marked by gray-shaded areas. The vertical gray-shaded band indicates the impulsive phase of the X3.4 flare.}
\label{fig:hv_methods_detailed}
\end{figure} 

In \href{fig:hv_methods_detailed}{Fig.~\ref{fig:hv_methods_detailed}}, we analyze the relative helicities computed by the different FV methods (see \href{tab:fv_methods_detailed}{Table~\ref{tab:fv_methods_detailed}} for the individual values). Though the DeVore methods deliver slightly smaller absolute values than the Coulomb methods, all methods are producing comparable values of $\Hdef$. Defining $\langle\Hdef\rangle$ the average value across the different FV estimations at a given time, one obtains $\langle\Hdef\rangle=[-2.79\pm0.20,-0.61\pm0.04,-1.32\pm0.08]\times10^{43}$~\mxmx\ for the December~11, 12, and 13 snapshots (see last column in \href{tab:nlff_modeling}{Table~\ref{tab:nlff_modeling}} and represented by black crosses in \href{fig:hv_methods_detailed}{Fig.~\ref{fig:hv_methods_detailed}(a)}). Here and in the following, mean values are given together with the corresponding standard deviation. The latter should not be considered to be a proper error on the mean, but rather a measure of the agreement between different FV methods of computation. For instance, in the case of $\Hdef$, the spread of solutions between all methods is 7.3\%, 6.8\%, and 5.7\% for the December~11, 12, and 13 snapshots, respectively. For the DeVore methods alone, the spread in $\Hdef$ is 0.2\%, 2.7\%, and 2.8\%, respectively.

From the point of view of the helicity decomposition in \href{eq:hj}{Eqs.~(\ref{eq:hj})} and \href{eq:hjp}{(\ref{eq:hjp})}, all FV computations result in a $\Hdef$ that is dominated by the volume-threading helicity ($\HdefJP$), with the current-carrying helicity ($\HdefJ$) comprising only $\lesssim10\%$ (compare \href{fig:hv_methods_detailed}{Fig.~\ref{fig:hv_methods_detailed}(b)} and \href{fig:hv_methods_detailed}{\ref{fig:hv_methods_detailed}(c)}, respectively). For the volume-threading helicity, we find average values of $\HdefJPm=[-2.52\pm0.2,-0.60\pm0.04,-1.28\pm0.08]\times10^{43}$~\mxmx\ for the December~11, 12, and 13 snapshots, corresponding to a spread of 8.0\%, 6.9\%, and 6.0\%, respectively. For the current-carrying helicity, we find average values of $\HdefJm=[-0.26\pm0.0004,-0.005\pm0.0005,-0.04\pm0.001]\times10^{43}$~\mxmx\ for the December~11, 12, and 13 snapshots, corresponding to a spread of 0.15\%, 9.6\%, and 2.8\%, respectively. 

\subsubsection{Intensive helicities}
\label{sss:hv_intensive}

\begin{figure}
\centerline{\includegraphics[width=0.9\columnwidth]{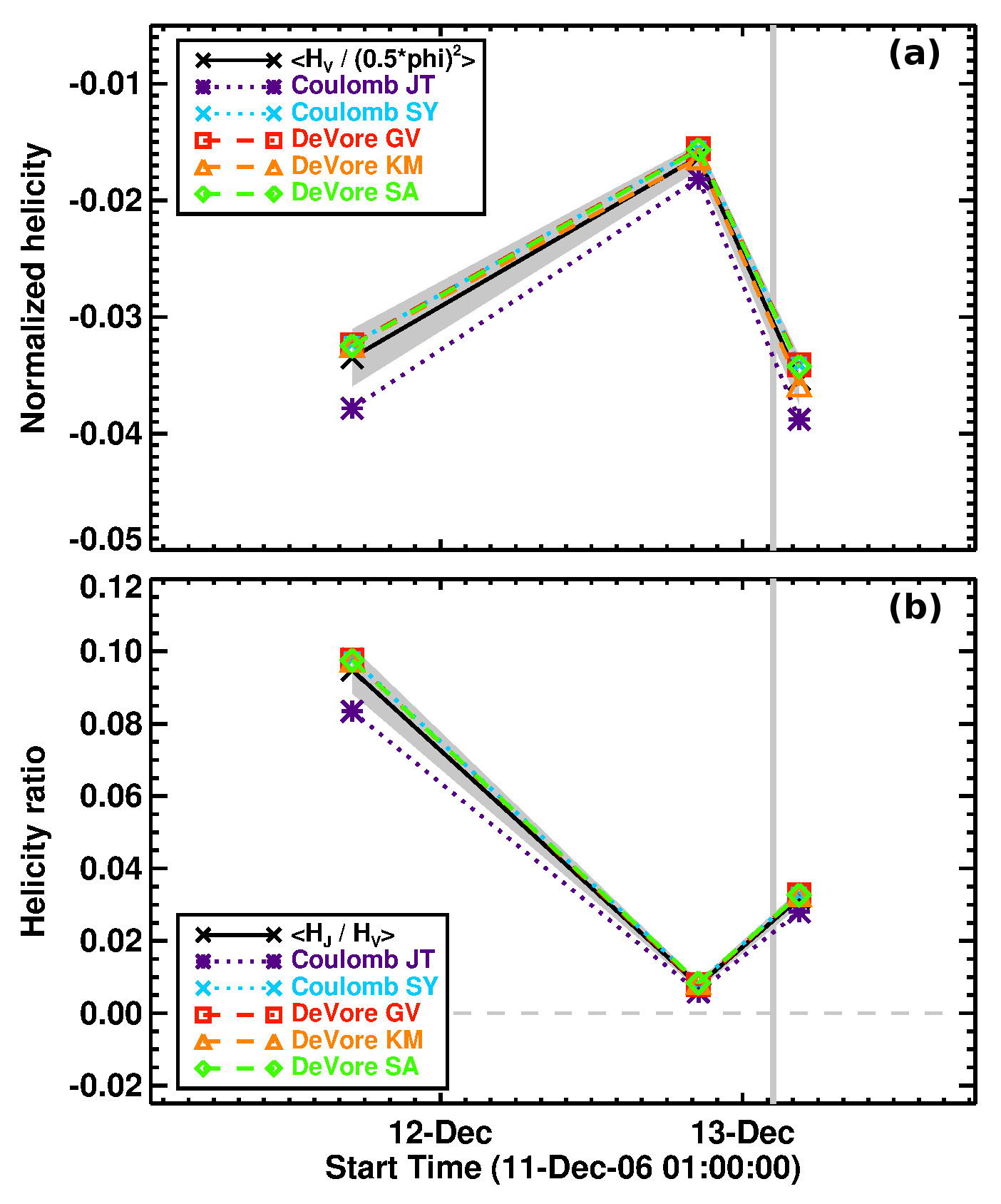}}
\caption{FV calculations of the intensive helicities for NOAA AR~10930 during 2006 December 11--13. (a) Normalized helicities ($\THdef$). (b) Helicity ratio ($\hjprime$). Layout as in \href{fig:hv_methods_detailed}{Fig.~\ref{fig:hv_methods_detailed}}.}
\label{fig:hv_methods_intensive_detailed}
\end{figure}

In \href{fig:hv_methods_intensive_detailed}{Fig.~\ref{fig:hv_methods_intensive_detailed}} we show the corresponding values for the normalized total helicity, $\THdef\equiv\Hdef/\phi^2$, and the helicity ratio, $\hjprime$ (cf.\ values listed in \href{tab:fv_methods_detailed}{Table~\ref{tab:fv_methods_detailed}}). These quantities are of particular interest as they harbor additional information on the non-potentiality of the coronal magnetic field.

All methods are basically producing the same trends for the normalized helicity, $\THdef$, (\href{fig:hv_methods_intensive_detailed}{Fig.~\ref{fig:hv_methods_intensive_detailed}(a)}). With the same precision as for $\Hdef$, the different method-based estimates deliver average values of $\langle\THdef\rangle=[-0.03\pm0.002,-0.02\pm0.001,-0.04\pm0.002]$ for the December~11, 12 and 13 snapshots, respectively. Taking all FV-based results into account, we find mean values of $\hjprimem=[0.09\pm0.006$, $0.01\pm0.001$, $0.03\pm0.002]$ for the December~11, 12 and 13 snapshots, agreeing to within 6.7\%, 14.1\% and 6.6\% (the DeVore methods alone to within 0.1\%, 3.6\% and 0.1\%), respectively. 

\subsection{Connectivity-based computations}
\label{ss:cb_results}

\subsubsection{Application to \CFITsc\ model lower boundary data}
\label{sss:cbff_results}

\begin{figure}
\centerline{\includegraphics[width=0.9\columnwidth]{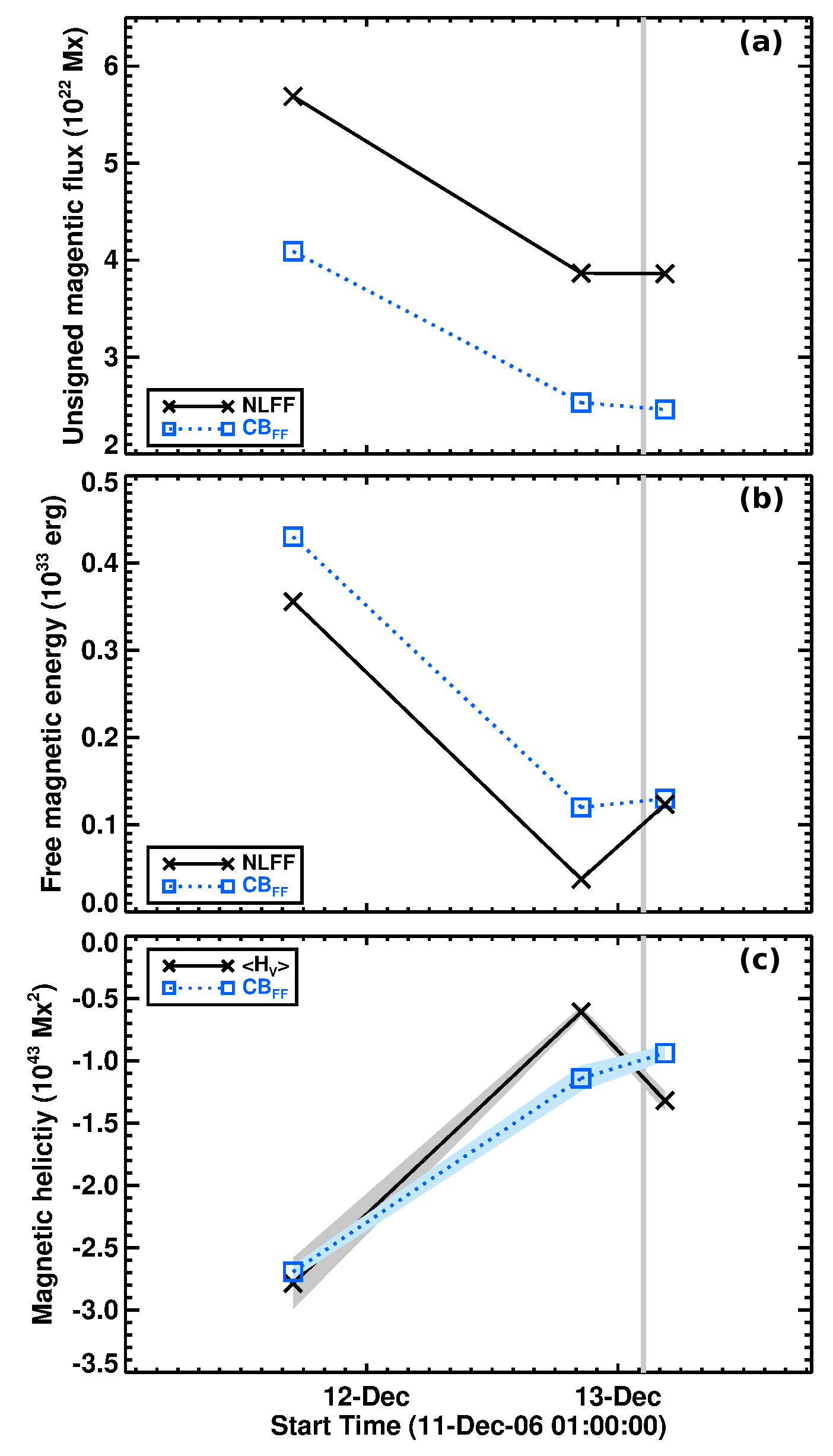}}
\caption{\CBFF\ calculations of the magnetic fluxes, energies and relative helicities for NOAA AR~10930 during 2006 December 11--13. (a) Total unsigned connected flux ($\phi_c$; blue squares) and \CFITsc\ lower boundary flux (black crosses). (b) \CBFF\ (squares) and \CFITsc\ (crosses) estimates of the free magnetic energy. (c) Mean FV ($\Hdefm$; crosses) and \CBFF\ (squares) relative helicities. Adjacent gray- and blue-shaded areas mark the corresponding uncertainties. The vertical gray-shaded band indicates the impulsive phase of the X3.4 flare.}
\label{fig:cbff_detailed}
\end{figure}

In \href{fig:cbff_detailed}{Fig.~\ref{fig:cbff_detailed}}, we show the physical quantities deduced from the \CBFF\ computations, i.e., from the application of the CB method to the \CFITsc\ lower boundary data. The respective values are listed in \href{tab:cbff_modeling}{Table~\ref{tab:cbff_modeling}} and are to be compared to the respective ones deduced from the \CFITsc\ coronal magnetic field models and subsequent FV helicity computations (\href{tab:nlff_modeling}{Table~\ref{tab:nlff_modeling}} in \href{ss:hv_results}{Sect.~\ref{ss:hv_results}}). Notable differences between the \CBFF\ and FV-based estimates, as discussed in the following, may primarily be due to the inherent property of the CB method to consider only a fraction of the total unsigned flux (via the connected flux $\phi_c$) of the supplied \CFITsc\ lower boundary data.

The total unsigned connected flux, $\phi_c$ of the CB method amounts to about 71\%, 65\% and 64\%, respectively, of the \CFITsc\ total unsigned fluxes for the December~11, 12 and 13 snapshots (compare blue squares and black crosses, respectively, in \href{fig:cbff_detailed}{Fig.~\ref{fig:cbff_detailed}(a)}). In other words, about 29\%, 35\% and 36\%, respectively, of the \CFITsc\ lower boundary flux is not considered by the CB computations, because the applied multi-polar partitioning assigns no corresponding closure within the considered computational domain.

The total magnetic energies deduced from the \CBFF\ method agree with the \CFITsc\ FV estimates to within a few percent of difference (compare corresponding values in \href{tab:nlff_modeling}{Tables~\ref{tab:nlff_modeling}} and \href{tab:cbff_modeling}{\ref{tab:cbff_modeling}}). In particular, $\ECB/\E=\left[0.99,1.01,0.96\right]$ for the December~11, 12 and 13 snapshots, respectively, while in case of the potential energy, $\EpCB/\Ep=0.96$ for all snapshots. The systematically lower \CBFF\ potential energy is due to using only a fraction of the total unsigned magnetic flux present. Differences are larger for the free magnetic energy, where $\EfreeCB/\Efree=\left[1.2,3.0,1.08\right]$, respectively (\href{fig:cbff_detailed}{Fig.~\ref{fig:cbff_detailed}(b)}). 

The \CBFF-based estimate of the total helicity, $\HdefCB$, matches the FV-based estimates only to some extent (\href{fig:cbff_detailed}{Fig.~\ref{fig:cbff_detailed}(c)}). The respective ratios $|\HdefCB|/|\Hdef|$ are $\left[0.96,1.87,0.71\right]$  for the three magnetogram snapshots.

\begin{table}
\caption{
Values of the unsigned connected magnetic flux ($\phi_c$), total ($E_{CB}$), potential ($E_{p,CB}$) and free ($E_{F,CB}=E_{CB}-E_{p,CB}$) magnetic energies, and coronal relative helicity, $\Hdef$, deduced from the application of the CB to the \CFITsc\ lower boundary data. Units of magnetic fluxes, magnetic energies and relative helicities are $10^{22}$\,\mx, $10^{33}$\,\erg\ and $10^{43}$\,\mxmx , respectively.}
\centering
\footnotesize
\strtable
\begin{tabular}{@{~}c c@{\quad} c@{\quad} c@{\quad} c@{\quad} c@{\quad} c@{\quad}}
\hline
Date \& Time & $\phi_c$ & $\ECB$ & $\EpCB$ & $\EfreeCB$ & $\HdefCB$ \\
\hline
 11 Dec 17:00~UT & 4.09 & 2.93 & 2.50 & 0.43 & -2.69$\pm$0.05\\
 12 Dec 20:30~UT & 2.53 & 1.90 & 1.78 & 0.12 & -1.14$\pm$0.10\\ 
 13 Dec 04:30~UT & 2.46 & 1.98 & 1.85 & 0.13 & -0.94$\pm$0.06\\
\hline
\label{tab:cbff_modeling}
\end{tabular}
\end{table}

\begin{figure}
\centerline{\includegraphics[width=0.93\columnwidth]{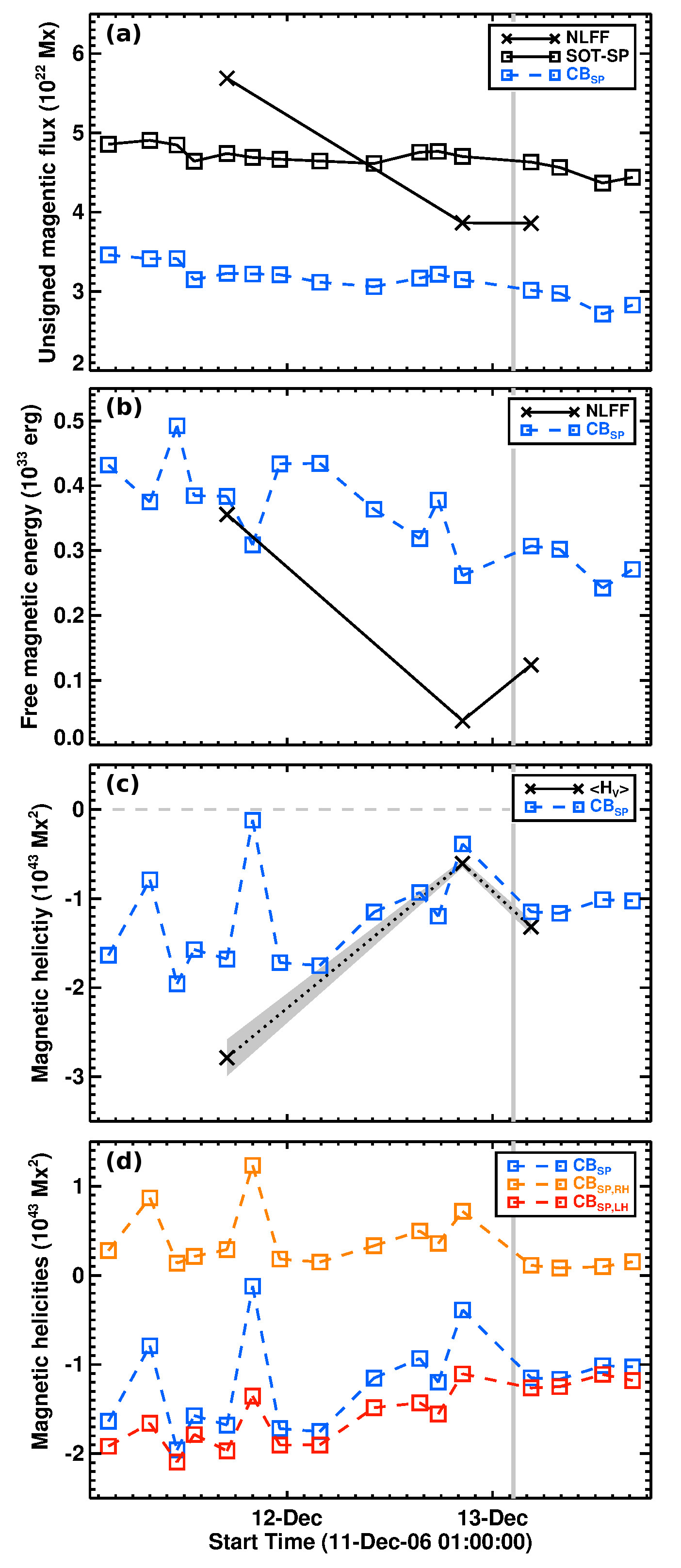}}
\caption{\CBSP\ calculations of the magnetic fluxes, energies and relative helicities for NOAA AR~10930 during 2006 December 11--13. (a) Total unsigned flux of SOT-SP data (black squares), total connected flux ($\phi_c$; blue squares), and \CFITsc\ lower boundary flux (black crosses). (b) \CBSP\ (squares) and \CFITsc\ (crosses) estimates of the free magnetic energy. (c) Mean FV ($\Hdefm$; crosses) and \CBSP\ (squares) relative helicities. (d) Contributions of $\HdefCBlh$ (red) and $\HdefCBrh$ (orange) to the \CBSP\ relative helicity (blue). The vertical gray-shaded band indicates the impulsive phase of the X3.4 flare.}
\label{fig:cbsp_detailed}
\end{figure} 

\subsubsection{Application to SOT-SP data}
\label{sss:cbsp_results}

We show the physical quantities deduced from the \CBSP\ computations in \href{fig:cbsp_detailed}{Fig.~\ref{fig:cbsp_detailed}} (for individual values see \href{tab:cbsp_detailed}{Table~\ref{tab:cbsp_detailed}}), in comparison to the FV estimates presented in \href{ss:results_cfit}{Sect.~\ref{ss:results_cfit}} and  \href{sss:hv_extensive}{Sect.~\ref{sss:hv_extensive}}. Let us clarify at this point that application of the CB method to 16 available SOT-SP vector magnetograms covers also the three time instances of the \CBFF\ application described in \href{sss:cbff_results}{Sect.~\ref{sss:cbff_results}}.  We remind the reader here that the SP data of this section have been prepared differently (for details see \href{ss:data_cb16}{Sect.~\ref{ss:data_cb16}}) than those for the FV (hence, \CBFF) computations, including  differences in linear size (field of view), spatial resolution, the azimuth disambiguation methodology and consideration of projection effects (for details see \href{tab:data}{Table~\ref{tab:data}}). Thus, notable differences between the SP- and FV-based estimates, as discussed in the following, may partly be due to differences in the underlying data preparation (see corresponding notes in \href{ss:results_cfit}{Sect.~\ref{ss:results_cfit}}), on top of the generally different approximation of magnetic connectivity in the coronal volume due to the CB-method induced magnetic flux partitioning.

The total unsigned magnetic flux, $\phi$ computed from $\Bz$ of the 16 Level-2 SOT-SP magnetograms is of the order $(4\;-\;5) \times10^{22}$\,\mx\ during the considered time period (i.e., between 11~December $\sim$03:10~UT and 13~December $\sim$16:21~UT). It shows a weak increase between 12~December $\sim$06:00~UT and $\sim$18:00~UT, followed by a more or less steady decrease until about 13~December 12:00~UT (black squares in \href{fig:cbsp_detailed}{Fig.~\ref{fig:cbsp_detailed}(a)}). The SOT-SP unsigned magnetic flux is lower (by $\approx$17\%) than that of the synthesized \CFITsc\ lower boundary for the December~11 snapshot, and about 20\% higher for the December~12 and 13 snapshots (compare black squares and black crosses, respectively, in \href{fig:cbsp_detailed}{Fig.~\ref{fig:cbsp_detailed}(a)}). The CB-based total connected flux, $\phi_c$, covers about 57\%, 81\% and 78\% of the \CFITsc\ lower boundary flux of the December~11, 12 and 13 snapshots, respectively (compare blue squares and black crosses in \href{fig:cbsp_detailed}{Fig.~\ref{fig:cbsp_detailed}(a)}).

The \CBSP\ computations for the total magnetic energy $\Etot$ amount to 74.4\%, 102.1\% and 93.3\% of the respective \CFITsc\ FV estimates for the December~11, 12 and 13 snapshots, while for the potential energy $\Ep$ \CBSP\ values are 69.8\%, 90.0\% and 83.4\% of the respective \CFITsc\ FV estimates (cf.\href{tab:nlff_modeling}{Tables~\ref{tab:nlff_modeling}} and \href{tab:cbsp_detailed}{\ref{tab:cbsp_detailed}}). As a consequence, the \CBSP\ and \CFITsc-based estimates of $\Efree$ agree for the December~11 snapshot (to within $\approx$7\%) while little agreement is found for the other two snapshots: the \CFITsc\ FV estimate of $\Efree$ amounts to 14.4\% and 40.2\%, respectively, for the December~12 and 13 snapshots.

At the corresponding time, the values of $\HdefCB$ are systematically smaller than $\Hdefm$: $|\HdefCB|$ amounts to 60.3\%, 63.8\% and 87.1\% of $\Hdefmabs$, for the December~11, 12 and 13 snapshots, respectively (blue symbols in \href{fig:cbsp_detailed}{Fig.~\ref{fig:cbsp_detailed}(c)}). Taking a closer look into the contributions to $\HdefCB$, we find a dominant left-handed contribution ($\HdefCBlh$), with a magnitude larger by a factor of $\sim$8, than the right-handed contribution ($\HdefCBrh$) (compare red and orange plus signs, respectively, in \href{fig:cbsp_detailed}{Fig.~\ref{fig:cbsp_detailed}(d)}).

One notices a couple of outlier points for magnetic helicity $\HdefCB$ in Figs.~\href{fig:cbsp_detailed}{\ref{fig:cbsp_detailed}(c)} and \href{fig:cbsp_detailed}{\ref{fig:cbsp_detailed}(d)}, particularly in the second and sixth points of the time series (08:00 and 20:00 UT on December 11). These have been judged to relate with local disambiguation issues that have resulted in opposite-sign helicity contributions from these localizations. These issues affect the magnetic free energy estimates (\href{fig:cbsp_detailed}{Fig.~\ref{fig:cbsp_detailed}(b)}), as well, but not as much as the relative helicity.

\subsection{Relative helicity flux}
\label{ss:h_flux}

\begin{figure}
\centering
\includegraphics[width=0.9\columnwidth]{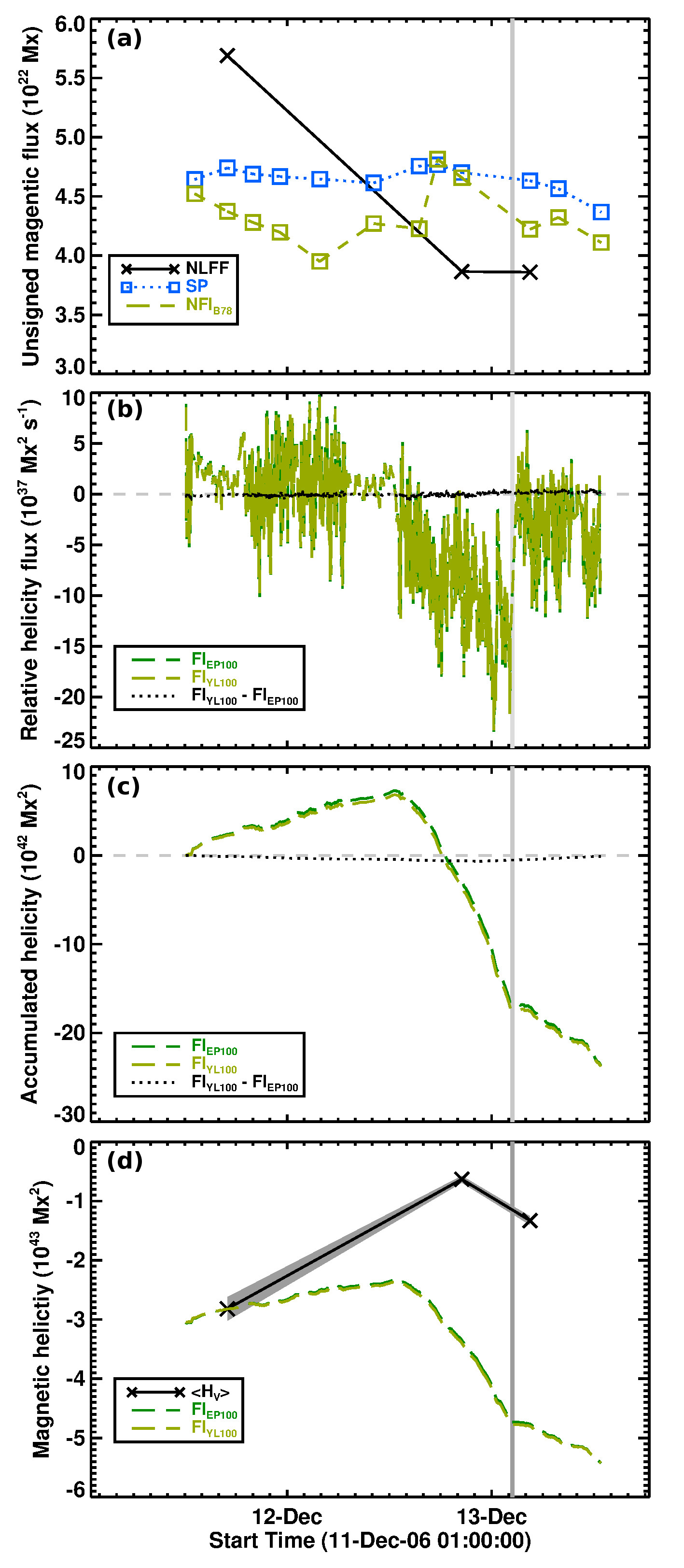}
\caption{Magnetic and (accumulated) helicity fluxes during the interval 11~December 12:14~UT -- 13~December 12:53~UT. (a) Unsigned magnetic fluxes computed from the calibrated NFI (green), SP (blue), and \CFITsc\ lower boundary (black) data. (b) Relative helicity fluxes computed from the calibrated NFI data. \FYL\ and \FEP\ computations are represented by dark and light green color, respectively. Their signed difference is indicated by a black curve. (c) Total accumulated helicity fluxes, $\Hacc$, and corresponding signed difference. (d) Mean coronal relative helicity, $\Hdefm$, (black crosses) and theoretical curves for $\Hacc$ when using the mean estimate of $\Hdefm$ on December~11 as a reference level. The vertical gray-shaded band indicates the impulsive phase of the X3.4 flare.
}
\label{fig:h_flx_detailed}
\end{figure}

In \href{fig:h_flx_detailed}{Fig.~\ref{fig:h_flx_detailed}(a)} we show the total unsigned magnetic flux, $\phi$, computed from the calibrated NFI $\Blos$ (green curve and squares), used as an input to the computational methods, \FYL\ and \FEP\ for relative helicity flux estimations. On overall, the unsigned fluxes of the calibrated NFI data agree with that computed from the SP data to within 15\%, and agree with the \CFITsc\ lower boundary fluxes to within 23\%, 20\%, and 9\%, for the 11~December 17:00~UT, 12~December 20:30~UT and 13~December 04:30~UT snapshots, respectively.

Temporal profiles of the calculated helicity flux are shown in \href{fig:h_flx_detailed}{Fig.~\ref{fig:h_flx_detailed}(b)} during the same three-day interval of December 11 -- 13. The helicity fluxes computed from the \FEP\ and \FYL\ methods, based on a threshold of 100~\gauss\ for $\Bz$ (represented by light and dark green curves, respectively, and labeled FI$_{{\rm EP}100}$ and FI$_{{\rm YL}100}$, respectively), result in very similar values (their signed difference is shown as a black dotted line), with an agreement to within $\approx$5\% (when considering all time instances when the unsigned helicity flux exceeds $1\times10^{37}$\,\mxmxs). Thus, the estimations of the relative helicity fluxes is largely consistent when computed by the \FEP\ and \FYL\ methods. Though not shown explicitly, we note here that the repetition of the \FYL\ helicity flux computation using a threshold of 20~\gauss\ for $\Bz$, yields helicity fluxes larger by $\approx$0.3\%, in comparison to the FI$_{{\rm YL}100}$ computations. This demonstrates that the helicity flux is mainly provided by the more intense magnetic polarities.

Overall, the period between 11~December $\sim$12:00~UT and 12~December $\sim$12:00~UT was characterized by a predominantly positive rate of photospheric magnetic helicity injection. The helicity injection rate appears to be rather constant around $\approx2\times10^{37}$\,\mxmxs. The second half of December~12 is characterized by a transition to strong negative values, roughly centered around $-10\times10^{37}$\,\mxmxs. This is followed by a transition to smaller negative values early on December~13, roughly around $-2.5\times10^{37}$\,\mxmxs.

By time integration of the helicity fluxes, without using a reference value for the coronal helicity as a starting value, we deduce the accumulated helicity $\Hacc$ as a function of time (\href{fig:h_flx_detailed}{Fig.~\ref{fig:h_flx_detailed}(c)}), and find different trends during distinct episodes. From both, the FI$_{{\rm YL}100}$ and FI$_{{\rm EP}100}$ computations (the signed difference between the two is shown as a black curve), we find that $\Hacc$ steadily increases, reaching peak values of $\approx7\times10^{42}$~\mxmx\ at 12~December $\sim$12:38~UT. Afterwards, $\Hacc$ decreases to negative values of $\approx-17\times10^{42}$~\mxmx\ at 13~December $\sim$02:13~UT (i.e., prior to flare onset), and further decreases to $\approx-24\times10^{42}$~\mxmx\ until 13~December $\sim$12:48~UT.

From the FI$_{{\rm YL}100}$ (FI$_{{\rm EP}100}$) computations, we estimate that a total of $-23.7\times10^{42}$\,\mxmx ($-23.6\times10^{42}$\,\mxmx) was injected through the photospheric boundary for the considered time period 11~December 12:09:20~UT -- 13~December 12:59:41~UT. We thus find that \FYL\ and \FEP\ estimation on $\Hacc$ are agreeing to within $\approx$8\%. For completeness, we note that using a 20~\gauss\ threshold instead of 100~\gauss\ only changes the precision of the \FYL\ computation of $\Hacc$ by $\approx$0.4\%.

Using our mean FV-based estimate for the total helicity on 11~December 17:00~UT as a reference, so that $\Hacc=\Hdefm$ at that time instant (see intersection of green dashed and black solid line in \href{fig:h_flx_detailed}{Fig.~\ref{fig:h_flx_detailed}(d)}), we construct a hypothetical time profile, $\Haccs$. The values obtained for $\Haccs$ (dark and light green solid lines) are qualitatively different from that of the FV-based $\Hdefm$ (black solid line). More precisely, $|\Haccs|$ exceeds $\Hdefmabs$ by a factor of $\sim$5.4 and $\sim$3.6 for the December~12 and ~13 snapshots (marked by black crosses), respectively. 

The FI methods, based on the analysis of more than 1150 magnetograms, naturally provides a more detailed description of the dynamic evolution of coronal helicity than the FV-based estimates (due to the coarse time resolution of the latter). Though we find $\Hdefmabs$ to be smallest for the December~12 snapshot, the cadence of the underlying \CFITsc\ models is too coarse as to allow us to assume with confidence that it represents a true peak in the time evolution of the coronal helicity. However, looking at the overall trends, we observe some qualitative agreement between the FV and the FI estimation: a decrease of negative helicity during the period spanning the second half of December~11 and the first half of December~12 and an increase in negative helicity after 12~December 20:30~UT. Quantitatively, we note a variation of $|\Delta\Haccs|\approx6.8\times10^{42}$~\mxmx\ ($7.2\times10^{42}$~\mxmx) from the \FYL\ and \FEP\ computations, respectively, between 11~December $\sim$12:09~UT and 12~December $\sim$12:38~UT, which is $\approx3$ times less than the variation of $\Hdefmabs$ during the same time span ($\approx2.1\times10^{43}$~\mxmx). Between 12~December 12:38~UT and 13~December 04:30~UT, $|\Haccs|$ shows a variation of $2.4\times10^{43}$~\mxmx\, while $\Hdefmabs$ varies by $\approx0.2\times10^{43}$~\mxmx, i.e., about 10 times less. Hence, while the FI and the FV methods show a partial agreement in terms of the time evolution of the coronal helicity, they quantitatively differ by several factors in the present application to observed data. Such large difference between the FI and FV methods was not observed in application to synthetic data \citep{2021SSRv..vol..pageP}.

\section{Discussion -- Method comparison}
\label{s:discussion}

In this study, we obtain the instantaneous coronal magnetic helicity budget from several FV helicity computation methods \citep[][]{2011SoPh..272..243T,2012SoPh..278..347V,2014SoPh..289.4453M} relying on various NLFF field extrapolations and its approximation from the CB method \citep[][]{2012ApJ...759....1G}, in comparison with the accumulated magnetic helicity derived from selected FI methods \citep[][]{2005A&A...439.1191P,2012ApJ...761..105L}. Based on high-quality (i.e., Level-2) photospheric \emph{Hinode}/SOT-SP vector magnetic field observations, in combination with \CFITsc\ magnetic field modeling \citep{2009ApJ...700L..88W,2011ApJ...728..112W}, we study the coronal magnetic energy and helicity of solar active region NOAA AR~10930 around an eruptive X3.4 flare (SOL2006-12-13T02:14). In the following, we discuss the main findings in regard to our main research objective, namely the cross-validation of different helicity computation methods.

\subsection{Comparison of FV results}
\label{ss:discussion_fv}

\href{ss:hv_results}{Sect.~\ref{ss:hv_results}} presents the results of FV methods when applied to real solar data. Given the high solenoidality of the NLFF fields used, assessed by the normalized fraction of the energy $\Ediv$ associated with magnetic monopoles (see \href{s:quality_append}{Appendix~\ref{s:quality_append}}), we do not expect a strong effect on helicity values because of such artifacts. The reader is also referred to dedicated analyses on solar applications by \cite{2019ApJ...880L...6T,2020A&A...643A.153T}. 

As already noted by \cite{2016SSRv..201..147V}, the accuracy of the FV helicity computed by different methods appears to be not directly related
to the accuracy of the vector potentials in reproducing the corresponding fields. More precisely, the \sJT\ method has a lower accuracy in solving for the vector potentials than the \sSY\ and the DeVore methods (cf.~\href{tab:fv_methods_detailed}{Table~\ref{tab:fv_methods_detailed}}), yet it delivers similar total ($\Hdef$) and decomposed helicities ($\HdefJP$ and $\HdefJ$). 

Overall, the results from the different FV methods differ by $\lesssim10.0\%$ from the common mean value, $\Hdefm$, when applied to the three \CFITsc\ models (\href{fig:hv_methods_detailed}{Fig.~\ref{fig:hv_methods_detailed}(a)}). The same is true for the decomposed helicities (\href{fig:hv_methods_detailed}{Fig.~\ref{fig:hv_methods_detailed}(b,c)}) and intensive (normalized) measures (\href{fig:hv_methods_intensive_detailed}{Fig.~\ref{fig:hv_methods_intensive_detailed}}). These findings thus verify and complement the results of \citet{2016SSRv..201..147V} and \citet{2019ApJ...880L...6T,2019ApJ...887...64T}, allowing us to assume with further confidence that FV methods provide consistent results on the local (i.e., active-region scale) coronal helicity content based on observational photospheric magnetic field data and the corresponding NLFF-extrapolated coronal magnetic fields.

\subsection{Comparison of FV and CB results}
\label{ss:discussion_cb}

In \href{ss:cb_results}{Sect.~\ref{ss:cb_results}}, the CB-based results have been compared to $\Hdefm$, the latter assumed to represent the ``ground-truth reference value'' of coronal helicity. This comparison between the CB and FV methods is by necessity limited to the estimated magnetic helicity and energy budgets since the CB method does not provide or utilize the vector potentials and reference fields. 

By design, the CB method considers only a fraction of the total unsigned flux (via the connected flux $\phi_c$) of the supplied input data (in the form of $\Bz$ at the \CFITsc\ lower boundary of the SP measurements). The extent to which magnetic information of the lower boundary is incorporated in the CB computations, however, does not seem to translate directly to a stronger or weaker agreement with FV-based values of $\Hdefm$. As an example, while $\phi_c$ is very similar for the December~12 and 13 snapshots for each method (both in terms of values and in terms of fraction to the total unsigned flux; see \href{fig:cbff_detailed}{Fig.~\ref{fig:cbff_detailed}(a)} and relevant discussion), the \CBFF\ helicity is $\sim 70$\% of $\Hdefm$ for the December 13 snapshot and $\sim 187$\% of $\Hdefm$ for the December 12 snapshot. On the contrary, despite a significantly higher $\phi_c$ on December 11, the helicity estimates still match to within $\approx$4\% (see \href{fig:cbff_detailed}{Fig.~\ref{fig:cbff_detailed}(c)}).

That implies, first, a non-linearity in the differences between CB- and FV-based helicity values even given a similar amount of $\phi_c$ and, second, that the match between CB- and FV-based helicities might not be necessarily better in case $\phi_c$ better matches the total unsigned magnetic flux. This may relate to the 'arch-like' magnetic-loops assumption of the CB method, which ignores intertwining of flux tubes in the corona because it does not require the essentially unknown full three-dimensional coronal field. This may include missing helicity contributions of both signs, thus giving rise to a nonlinear effect in the comparison.

Another nonlinear effect appears in the free magnetic energy, in which the contribution by missing braided coronal connections is always positive. Perhaps surprisingly, the CB-based estimates of the free energy are systematically larger than those based on the \CFITsc\ models, by factors of 1.2, 3.0 and 1.08, respectively (\href{fig:cbff_detailed}{Fig.~\ref{fig:cbff_detailed}(b)}). While it is clear that $\EfreeCB$ is an underestimation of the true magnetic free energy in the corona, its systematic excess of $\Efree$ values may imply that the NLFF field extrapolations give rise to relatively smooth magnetic fields, closer to a potential-field solution than the true field.

The above said, the \CBFF\ and \CFITsc\ results in both, free energy and helicity, are not more than a factor of three different (a factor of $\lesssim$2 for the helicity), agree in helicity sign, and provide a roughly similar evolution of the studied NOAA AR 10930, in showing a decrease of the magnetic free energy and helicity budgets between December 11 and December 13. In order words, both describe a gradual relaxation of the magnetic structure in the active region. We elaborate  on this physical evolution in more detail in \href{s:discussion_extended}{Sect.~\ref{s:discussion_extended}}.

Overall agreements regarding FV- and CB-based estimates of the instantaneous coronal energy and helicity budgets can be found by comparison of other independently performed analyses. For instance, \cite{2013ApJ...772..115T} and \cite{2019ApJ...887...64T} independently studied the long-term evolution of AR~11158, showing an overall agreement of the time profiles deduced from application of the CB and a FV method, respectively (see their Figures 2e and 3b, respectively). The fact that the absolute \CBSP-based estimates are a factor of two higher than corresponding FV-based estimates may be due to several reasons, including differences in the spatial resolution of the underlying magnetic vector data and the considered FOV. This said, the overall increasing trends of helicity and free energy can be found in both studies. In another application by \cite{2016ApJ...817...14P}, the coronal helicity budget timely around a pair of X-class flares triggered in AR~11429 was studied, revealing that the CB and a FV method agree in the sense of helicity recovered (a predominantly left-handed structure), with a factor of $\sim$2 difference in helicity amplitudes (see their Table 2).

Our analysis of the CB-based helicities in Sect.~\href{sss:cbsp_results}{\ref{sss:cbsp_results}} also shows differences between the same method (CB) when applied to SP data differing in spatial resolution, field of view (yet encompassing the essential central part of the active region), and particular steps taken in data preparation (disambiguation method and/or additional embedding in case of the \CBFF\ computations). While the \CBFF\ and \CBSP\ results agree qualitatively in terms of trends describing the physical evolution of the active region, it is difficult to disentangle the different quantitative effects without additional tests. This testing is left for a dedicated future work.

\subsection{Comparison of FV, CB and FI results}
\label{ss:discussion_fi}

From application of the two tested FI methods to a high-cadence time series of NFI $\Blos$ magnetic field data (\href{ss:h_flux}{Sect.~\ref{ss:h_flux}}), we find strong agreement between the \FYL\ and \FEP\ methods, to within ($\approx$8\%) $\approx$5\% regarding the (accumulated) helicity flux, when using the same thresholds on the level above which values of $\Blos$ are considered for FI computations. This is fully consistent with the results of \cite{2021SSRv..vol..pageP}, where a similarly good agreement between the \FYL\ and \FEP\ methods was found when applied to different synthetic data produced by 3D numerical simulations of solar-like events. Furthermore, when varying the threshold of $\Blos$, we find the \FYL-based estimates of the helicity flux and accumulated helicity to agree within $\approx$0.3\% and $\approx$0.4\%, respectively, indicating the particular threshold used not to play a crucial role, i.e., suggesting that it is mostly the intense magnetic field area that contributes to the helicity budget.

Taking the FV-based mean estimate of the coronal helicity budget for 11~December 17:00~UT as a reference, i.e., $\Hacc=\Hdefm$ at that time instant, the relative helicity accumulation, $\Haccs$, suggests a decrease of the coronal helicity budget during the first half of the pre-eruption phase (between 11~December 17:00~UT and 12~December 12:38~UT) of about $|\Delta\Haccs|=0.5\text{--}0.6\times10^{43}$~\mxmx (\href{fig:h_flx_detailed}{Fig.~\ref{fig:h_flx_detailed}(d)}). This is quite consistent with the decrease in coronal helicity during the same period as estimated from the \CBSP\ computations ($|\Delta\HdefCB|\approx0.7\times10^{43}$~\mxmx; compare \href{fig:cbsp_detailed}{Fig.~\ref{fig:cbsp_detailed}(c)}), but is less consistent with the overall trend seen in the FV- and \CBFF-based total helicity, the latter suggesting the corresponding decrease in coronal helicity to be larger by a factor of $\approx$3 and $\approx$2, respectively (compare \href{fig:hv_methods_detailed}{Figures~\ref{fig:hv_methods_detailed}(a)} and \href{fig:cbff_detailed}{\ref{fig:cbff_detailed}(c)}), respectively).

In this study, however, the absolute values obtained for $\Haccs$ (constructed using $\Hdefm$ at 11~December 17:00~UT as a reference level) are quite different from FV-based mean estimate, $\Hdefmabs$, the latter recovering only $\approx$19\% and $\approx$28\% of $|\Haccs|$ for the December~12 and 13 snapshots, respectively (see \href{fig:h_flx_detailed}{Fig.~\ref{fig:h_flx_detailed}(d)}). This weak agreement (to within $\lesssim30\%$ only, also observed in context with the corresponding CB-based estimates) is in line with the lack of correspondence between FV- and FI-based helicity estimates found in other applications \citep[see e.g.,][]{2008ApJ...682L.133Z,2010ApJ...720.1102P}, but contrasts the findings based on controlled experiments, carried out by \cite{2021SSRv..vol..pageP}, which showed a good correspondence (at least to within $80\%$). 

In principle, difference between FV and FI measurements are expected for flare-induced changes to the coronal helicity budget because, in contrast to FV computations, the FI methods are unlikely capable of tracking the amount of helicity carried away by a CME and the associated reorganisation of helicity within the coronal domain. This was also supported by \cite{2021SSRv..vol..pageP}, who reported strong deviations of the different helicity measures during the eruptive phase. Nevertheless, in our study, the FI-based time evolution of the coronal helicity between the pre-flare (at 12~December 20:30~UT) and post-flare (at 13~December 04:30~UT) corona appears consistent, indicating an increase of coronal helicity. We find, in particular, $|\Delta\Haccs|\simeq1.4\times10^{43}$~\mxmx, in comparison to $|\Delta\Hdefm|\simeq0.71\times10^{43}$~\mxmx, i.e., an agreement to within $\approx$50\%.

Overall agreements regarding CB- and FI-based estimates of the instantaneous helicity budget have been demonstrated in \cite{2016ApJ...817...14P} in their analysis of AR~11429 on 2012 March 7. In that study, the two helicity calculation methods agreed in a predominant left-handed (negative) helicity in the AR, but with a magnitude differing by a factor of $\sim$8 (see their Table 2). Surprisingly, in \cite{2016ApJ...817...14P} the FI method gave by far the largest helicity estimate, with $\Hacc$ also by a factor of $\sim$4 larger than a corresponding FV computation.

\section{Extended discussion -- Physical interpretation}
\label{s:discussion_extended}

Our rather extensive analysis of NOAA AR~10930 affords us a picture of the complicated events that preceded the eruptive X3.4 flare, starting about 1.5 days prior to the event. In the following, we interpret our main findings in regards of the SOL2006-12-13T02:14X3.4 eruption, as well as of the active region evolution that led to it, and place them into context with existing literature. 

\subsection{Pre-flare evolution}
\label{ss:discussion_preflare}

We find that the interval 11~December $\sim$12:00~UT -- 12~December $\sim$13:00~UT was characterized by a predominantly right-handed (positive) rate of magnetic helicity injection through the photosphere (\href{fig:h_flx_detailed}{Fig.~\ref{fig:h_flx_detailed}(b)}). This resulted in an accumulation of $\Hacc\simeq6.8\times10^{42}$\,\mxmx\ of positive helicity in the corona (\href{fig:h_flx_detailed}{Fig.~\ref{fig:h_flx_detailed}(c)}). Afterwards, the rate of helicity injection transited to strong negative values, persisting until just before the time of the X-class flare on December~13 and resulting in a total of $\Hacc\approx17\times10^{42}$\,\mxmx\ in left-handed (negative) helicity at 20:30~UT on December 12. Excluding the FI-based estimates during the nominal flare duration, we find a further increase of the total accumulated coronal helicity budget until the end of the investigated time period at 13~December $\sim$12:00~UT, amounting to a total of $\Hacc=-23.7\times10^{42}$~\mxmx\ during the entire analysis interval.

Notice that our estimates of the (accumulated) helicity flux are roughly an order of magnitude larger than those in earlier studies \citep[e.g.,][]{2008ApJ...682L.133Z,2010ApJ...720.1102P}.
From those studies, and taking our estimate of $\Hdefm$ as a reference, one would conclude that $\Hacc$ accounted for only a minor contribution to the coronal helicity budget. Instead, we find in this study that $\Hacc$ contributes markedly and evolves only partly consistently in time with the FV-based estimates. We attribute the discrepancy between our results and those published earlier to the challenge of proper data calibration, i.e., the quality of the photospheric magnetic field data used to carry out the helicity flux computations. In short, only when omitting any calibration of the NFI data, we are able to reproduce the (accumulated) helicity fluxes of, e.g., \cite{2008ApJ...682L.133Z} and \cite{2010ApJ...720.1102P}. Given the strong difference between the results obtained with or without calibration (cf.\ \href{ss:calib_append}{Appendix~\ref{ss:calib_append}}), our study points to the care requested when handling the input magnetic field data in order to properly use the FI methods with observed data.

From our analysis of individual contributions to volumetric estimates (\href{fig:hv_methods_detailed}{Fig.~\ref{fig:hv_methods_detailed}}), a clear dominance of the volume-threading helicity ($\HdefJP$) is recovered, being about an order of magnitude larger than the current-carrying helicity ($\HdefJ$). Corresponding dominant contributions of $\HdefJP$ are known from earlier simulation-based \citep[e.g.,][]{2017A&A...601A.125P,2018ApJ...863...41Z,2018ApJ...865...52L} and observation-based \citep[][]{2018ApJ...855L..16J,2019A&A...628A..50M,2019ApJ...887...64T,2019A&A...628A.114P} works. Yet puzzling are our estimates of the helicity ratio ($\hjprime$; \href{fig:hv_methods_intensive_detailed}{Fig.~\ref{fig:hv_methods_intensive_detailed}(b)}). From observational studies of individual ARs prolific in eruptive X-class flares, pre-flare peak values of $\gtrsim$0.15 were found \citep[e.g.,][]{2019A&A...628A..50M,2019ApJ...887...64T}. The comparative recent work by \cite{2021ApJ...vol...ppG}, in which ten different ARs are studied, places these values to an extreme, with CME-productive ARs showing characteristic pre-flare values of $\gtrsim$0.1. In sharp contrast, we find a corresponding mean FV-based pre-flare estimate of $<$0.1 (\href{fig:hv_methods_intensive_detailed}{Fig.~\ref{fig:hv_methods_intensive_detailed}(b)}), which might be related to our NLFF models being more potential (with $\Efree/\Etot\lesssim0.1$) than in earlier studies of CME-productive ARs.

The time evolution of $\Hdefm$ suggests a decrease of the coronal helicity budget during December~12. Quantitatively, we find the total helicity to decrease between 11~December 17:00~UT and 12~December 20:30~UT by $2.2\times10^{42}$\,\mxmx\ (see \href{fig:hv_methods_detailed}{Fig.~\ref{fig:hv_methods_detailed}}(a)), in overall agreement with the decrease in coronal helicity evaluated by \cite{2010ApJ...720.1102P} (see their Fig.~1) and \cite{2012ApJ...759....1G} (see their Fig.~7a) during the same time period. From the \CBSP\ computations (\href{fig:cbsp_detailed}{Fig.~\ref{fig:cbsp_detailed}(d)}), one notices a dominant left-handed contribution ($\HdefCBlh$) decreasing during the same period, consistent with the assumption of an magnetic configuration of negative overall helicity. A co-temporal weak increase of the corresponding right-handed contribution ($\HdefCBrh$), however, suggests the emergence of an oppositely helical (i.e., right-handed) magnetic structure, consistent with our finding of a positive helicity flux discussed above.

Overall, the above described trends support a scenario of a right-handed structure emerging into a pre-existing, predominantly left-handed magnetic configuration during December~11 and the first half of December~12. This is consistent with the NLFF model-based findings of \cite{2012ApJ...760...17I} who showed that the active-region magnetic field was predominantly negatively twisted about one day prior to the X-class flare, as well as the formation of positively twisted field near the polarity inversion line prior to flare onset. Consistently, we recover a positively sheared arcade from our \CFITsc\ models. A system of strong electric currents is found in the arch filament system on December~11 17:00~UT and an associated low-lying sheared arcade connecting the two main sunspots on December~12 at 20:30~UT (\href{fig:sample_fls}{Fig.~\ref{fig:sample_fls}(b)}). 

\subsection{Pre- and post-flare conditions in comparison}

A complicated, challenging picture also appears in comparing pre- and post-flare configurations in regards to the major, GOES X3.4 flare in the active region over the studied interval.

First, we interpret a sign reversal in the photospheric helicity flux during the impulsive phase of the flare to be nonphysical, contrary to earlier studies \cite[e.g.,][]{2008ApJ...682L.133Z,2010ApJ...720.1102P,2011ApJ...740...19R}. In those studies, it was suggested to represent a signature of the rapid emergence of a magnetic structure of opposite handedness, possibly responsible for the triggering of the flare. In our study, however, we present support that helicity flux estimates during this particular flare lack realism, and excluded those during the nominal flare duration from analysis (for details see \href{ss:disclosure_append}{Appendix~\ref{ss:disclosure_append}}). Consequently, we question the interpretation of a sudden and impulsive helicity injection as the trigger of the X3.4 flare, and refer to \cite{2007ApJ...671..955L} and \cite{2018NatCo...9...46X} for the discussion of better observed, and more credible, flare-related changes.

Second, from our FV and \CBSP\ computations, we find an increase in the coronal helicity between 12~December 20:30~UT and 13~December 04:30~UT of $|\Delta\Hdefm|\simeq0.71\times10^{43}$~\mxmx\ and $\simeq$$0.76\times10^{43}$~\mxmx, respectively, in line with earlier works \citep[e.g.,][]{2010ApJ...720.1102P,2012ApJ...759....1G}. Since our FV-based decomposition of the total helicity allows it, we find the differences between the pre-flare and post-flare snapshots to be more pronounced in the volume-threading ($\HdefJP$) than in the current-carrying ($\HdefJ$) helicities ($\approx$11\% vs.\ $\approx$6\%, respectively, of the pre-flare value of $\Hdefm$), and more pronounced in the right-handed ($\HdefCBrh$) than the left-handed ($\HdefCBlh$) contribution to $\HdefCB$ ($\approx56\%$ vs.\ $\approx41\%$ of the pre-flare $\HdefCB$, respectively). Thus, we may assume with relative confidence that $\HdefJ$ (showing a flare-related increase, as does $\HdefCBlh$) is dominated by the core (left-handed) field in the active region, left behind after the ejection of a previously emerged right-handed structure.

Third, based on our \CFITsc\ magnetic field models, we find the free magnetic energy ($\Efree$) to be higher for the post-flare configuration (\href{tab:nlff_modeling}{Table~\ref{tab:nlff_modeling}}). From our volumetric estimates on 12~December 20:30~UT and 13~December 04:30~UT, we quantify the corresponding increase as to be $\Delta\Efree\approx1\times10^{32}$~erg. Also the fraction of $\Efree$ compared to the total energy is higher in the post-flare corona ($\approx$6\%, compared to $\approx$2\% for the pre-flare corona). This increasing trend is in line with the results obtained from 13 out of 14 NLFF solutions compared in \cite{2008ApJ...675.1637S}, and is in line with the findings of \cite{2008ApJ...676L..81J} regarding the increase of magnetic shear in the course of the flare.

Consistent increasing trends are found from the \CBFF\ (\href{tab:cbff_modeling}{Table~\ref{tab:cbff_modeling}} and \href{fig:cbff_detailed}{Fig.~\ref{fig:cbff_detailed}(b)}) and \CBSP\ (\href{tab:cbsp_detailed}{Table~\ref{tab:cbsp_detailed}} and \href{fig:cbsp_detailed}{Fig.~\ref{fig:cbsp_detailed}(b)}) results, suggesting however an increase of $\Efree$ by a factor of eight and two lower than the \CFITsc-based estimates, respectively. Regardless, these findings contradict those of other studies. The ${\rm Wh}_{\rm pp}^+$ NLFF modeling reported in \cite{2008ApJ...675.1637S}, suggests a flare-related decrease of $\Delta\Efree\approx3\times10^{32}$\,\erg, in loose agreement (about an order of magnitude higher) with the corresponding estimate of \cite{2008ApJ...679.1629G}. Being necessarily related to the differing pre-flare magnetic topology recovered from our \CFITsc\ modeling, the discrepancy regarding the recovered time evolution of the coronal magnetic energies may again be attributed to the overall uncertainties and ambiguity of NLFF modeling. 

\section{Summary}
\label{s:summary}

The study and analysis presented herein serves the primary purpose of cross-validating different calculation methods of the relative magnetic helicity in the well-studied AR~10930 around the time of an eruptive major flare (SOL2006-12-13T02:14X3.4). It is part of a series of ISSI-supported studies devoted to comparisons between the results of different helicity calculation methods \citep{2016SSRv..201..147V, 2017ApJ...840...40G,2021SSRv..vol..pageP} and is the first study of the series employing solar observations.

To the above objective, we used the following helicity calculation methods:

\small
\begin{itemize}
  \item Five different finite volume (FV) methods (cf.\ \href{ss:fv_methods}{Sect.~\ref{ss:fv_methods}}), relying on the classical volume-integral magnetic helicity formula, applied to a series of three NLFF field extrapolations (\CFITsc; cf.\ \href{ss:data_fv}{Sect.~\ref{ss:data_fv}}). The \CFITsc\ modeling used a synthetic photospheric boundary, constructed from \emph{Hinode} SOT-SP vector magnetograms and \emph{SOHO}/MDI LOS magnetograms (\href{ss:data_nlff}{Sect.~\ref{ss:data_nlff}}).
  \item The connectivity-based (CB) method, relying on a partitioning of photospheric magnetic flux distributions (\href{ss:dt_methods}{Sect.~\ref{ss:dt_methods}}), applied to two different sets of photospheric boundary data: once to the \CFITsc\ lower boundary vector magnetic field (\CBFF), and once to a time series of 16 Level-2 SOT-SP vector magnetograms (\CBSP).
  \item Two different helicity-flux integration (FI) methods (\href{ss:fi_methods}{Sect.~\ref{ss:fi_methods}}), relying on a high-cadence time series of 1150 \emph{Hinode} SOT-NFI LOS magnetograms (\FYL\ and \FEP\ methods). 
\end{itemize}
\normalsize

The FV and CB methods provide instantaneous budgets of the magnetic free energy ($\Efree$) and relative helicity ($\Hdef$) in the active-region corona, while the FI methods provide the helicity injection rate through the photosphere and an accumulated (i.e., time-integrated) helicity ($\Hacc$) thereof. 

In regards of our main research objective, namely the cross-validation of different methods, we found a number of promising aspects:

\small
\begin{itemize}
  \item[{\sc (i)}] A close correspondence between FV estimates, both in extensive and intensive estimates, with an agreement to within a few percent.
  \item[({\sc ii})] Agreement on the dominant (left-handed) helicity in the AR as deduced from the FV and CB methods.\\
  Overall agreement between FV- and CB-based estimates regarding recovered time trends, deemed as remarkable given the very different settings of the methods: the CB method only models the coronal magnetic connectivity while the FV methods requires it as an explicit input.
  \item[({\sc iii})] A close correspondence between FI estimates, with an agreement to within a few percent.
  \item[({\sc iv})] Overall agreement between FV- and FI-based estimates regarding the predominant sign and magnitude of the coronal helicity. This is also deemed as remarkable, given that the FI method only capture the flux of helicity supplied to the corona via the photosphere.
\end{itemize}
\normalsize

In terms of the second objective, namely the interpretation of the active region evolution that led to the SOL2006-12-13T02:14X3.4 eruption, we found an overall decreasing free magnetic energy and relative magnetic helicity during 2006 December 11 -- 12, and increased values for the post-flare corona on December~13. All FV, \CBFF\ and \CBSP\ results basically corroborate this picture, with the \CBSP\ method further implying the possible expulsion of an oppositely helical (i.e., right-handed) structure in the course of the eruption that was previously embedded (and, possibly, emerged during the previous 24 hours) into a predominantly left-handed magnetic configuration.

In this analysis, furthermore, we encountered and identified a number of significant caveats: 

\small
\begin{itemize}
  \item[(a)] Our \CFITsc\ model results are significantly different than the best-performing (${\rm Wh}_{\rm pp}^+$) NLFF model in the community-supported study of \cite{2008ApJ...675.1637S}, even to the point of leading to different physical interpretations. Our results allow interpretations in line with 13 out of 14 NLFF models analyzed in that work \citep[and also in line with][]{2008ApJ...676L..81J}, pointing at the long-known uncertainties and ambiguities of NLFF modeling.
  \item[(b)] Our \CFITsc\ models do not show a coronal flux rope being present in the pre-flare corona (a highly sheared arcade instead). Studies advocating for the pre-flare existence of flux ropes exist \citep[e.g.,][]{gibson_etal06}, while recent reviews advocate for an evolutionary course from an initial sheared magnetic arcade to a magnetic flux rope during the eruption \citep[][and references therein]{2019RSPTA.37780094G,2020SSRv..216..131P}. 
  \item[(c)] Our interpretation of a possible ejection of an oppositely helical magnetic structure stands in agreement with some modeling works \citep[e.g.,][]{2012ApJ...760...17I} but disagrees with others \citep[e.g.,][]{2016ApJ...824...93F} that modeled an ejected helical structure with like (i.e., left-) handedness to the prevailing helicity sense in the active region. 
  \item[(d)] For the FV and CB methods, photospheric boundary conditions are crucial to the quantitative results. Nevertheless, the match of the FV-based estimates from the CB method is apparently unrelated to the fraction of the true magnetic flux considered.
  \item[(e)] \CBFF\ and \CBSP\ results are significantly different, relying on \CFITsc\ lower boundary and original SP magnetograms, respectively. 
  \item[(f)] FI computations based on high-cadence, high-quality photospheric vector magnetograms, allow a detailed and realistic account of the helicity buildup rate in an active region, hence they critically depend on data calibration.
  \item[(g)] Spurious signals in the FI results (showing a fast transition to helicity injection of opposite sign) during the nominal duration of the X3.4 flare, might be wrongly interpreted as representing an observational signature of a right-handed magnetic structure having impulsively emerged. 
  \item[(h)] FI computations need a reference point for the starting helicity budget to complete the evolutionary picture. Even with such a reference level at hand (e.g., via a FV-based estimate), significant discrepancies between the deduced helicity budgets may be found (up to $\approx70\%$ in this study), lacking a sound, well-justified explanation to date. 
  \item[(i)] In terms of the free magnetic energy, \CBFF\ results give consistently higher values than \CFITsc\ results, albeit by reasonable factors 1--3 (compare Tables \ref{tab:nlff_modeling} and \ref{tab:cbff_modeling}), despite representing a minimum coronal free energy by construction of the CB method. This points to a potential underestimation of the free energy in the corona by NLFF field extrapolations.
\end{itemize}

\normalsize

In brief, this study highlights the intricacies and difficulties of interpreting a complexity-ridden solar eruption by means of a quantitative data analysis of its host active region. While the X3.4-flare related eruption must have undoubtedly resulted in magnetic energy release and helicity expulsion, our results suggest enhanced respective budgets at a time shortly afterwards. One might be quick to discard these results as counter-intuitive or less credible, if only a single method to estimate the coronal energy and helicity budgets were used. The independent overall agreement by different methods in this study, however, may nudge one toward a slightly more complicated picture, namely, one of competing tendencies in the active region. Regardless how large an eruptive flare, it only releases a relatively small fraction of the free energy in the region (up to $\sim10$\%) and up to $\sim$30-40\% of its helicity \citep[e.g.,][]{2003ApJ...594.1033N,2008JGRA..113.9103G,2014SoPh..289.4453M}. Hence, tendencies of buildup or decay may well be stronger than the respective eruption budgets imposing the need for a wider investigation of active region evolution and dynamics around eruptions, rather than a mere focus on the eruptions themselves.

\acknowledgments
{\footnotesize
We thank the referee for the positive evaluation of our manuscript and for the comments and suggestions to optimize it. J.~K.~Thalmann acknowledges support from the Austrian Science Fund (FWF): P31413-N27. M.\,G.\ acknowledges support by the past SoME-UFo Marie Curie Fellowship of the European Commission (grant agreement no. 268245), in the framework of which the connectivity based method was developed and published. E.\,P.\ acknowledges support of the French Agence Nationale pour la Recherche through the HELISOL project ANR-15-CE31-0001. G.\,V.\ acknowledges support of the Leverhulme Trust Research Project Grant 2014-051. S.\,A.\ is financially supported by RFBR grant 18-29-21016\_mk and by the Ministry of Science and Higher Education of the Russian Federation. Y.\,G.\ is supported by NSFC grants 11773016, 11733003, and 11533005. F.\,C.\ is supported by the Fundamental Research Funds for the Central Universities under grant 0201-14380041. S.\,Y.\ acknowledges support by grants 11427901, 10921303, 11673033, U1731113, 11611530679, and 11573037 of the National Natural Science Foundation of China and grants no. XDB09040200, XDA04061002, XDA15010700 of the Strategic Priority Research Program of Chinese Academy of Sciences and the Youth Innovation Promotion Association of CAS (2019059). A.\,M.\ was supported by Australian Research Council Discovery Project DP160102932. \emph{Hinode} is a Japanese mission developed and launched by ISAS/JAXA, with NAOJ as domestic partner and NASA and STFC (UK) as international partners. It is operated by these agencies in co-operation with ESA and NSC (Norway). This article profited from discussions during the meetings of the ISSI International Team {\it Magnetic Helicity estimations in models and observations of the solar magnetic field} (\url{http://www.issibern.ch/teams/magnetichelicity/}).
}

\clearpage
\appendix

\section{Data sources and preparation}
\label{s:data_append}
\setcounter{figure}{0}
\renewcommand{\thefigure}{A\arabic{figure}}
\setcounter{table}{0}
\renewcommand{\thetable}{A\arabic{table}}

We summarize the sources and processing particular data used for NLFF modeling and/or helicity computations in \href{tab:data}{Table~\ref{tab:data}}, including the time (range) for which data was acquired (first column), the number of snapshots considered (second column), the particular data source used (third column), as well as the plate scale of the resulting data product (fifth column). If applied, the method used for disambiguation of the magnetic field azimuth is listed in the fourth column. The remaining columns point at the places in the manuscript where the corresponding FOVs are visualized (sixth column) and explained in detail (last column).

\begin{table}[ht]
\caption{Data sources and preparation for \CFITsc\ modeling and the application of the different helicity computation methods. Indicated from left to right are the instances or time range, where applicable, of data coverage, the number of snapshots used within the covered time range, the data source, disambiguation method (where applicable; otherwise a cross '$\times$' is used), the plate scale, the indication of the covered area on the solar disk in Fig.~\href{fig:fov}{\ref{fig:fov}} (for data other than on December 11, a cross '$\times$' is used) and the location of detailed data description in the main document.}
\centering
\footnotesize
\strtable
\begin{tabular}{| c@{\quad} c@{\quad} c@{\quad} c@{\quad} c@{\quad} c@{\quad}c@{\quad}|}
\hline
  Time (range) & No.\ of & Data & Disambiguation & Plate scale & Covered area & Detailed\\
  (UT) & snapshots & source & method & (arcsec) & as outlined in \href{fig:fov}{Fig.~\ref{fig:fov}} & description\\
 \hline
 \multicolumn{7}{|c|}{\bf \CFITsc\ NLFF modeling}\\
  Dec~11 17:00 & 1 & SOT-SP$^a$ & NPFC$^b$ & 0.66 & magenta outline & \href{ss:data_nlff}{Sect.~\ref{ss:data_nlff}}\\
  Dec~12 20:30 & 1 & SOT-SP$^a$ & ME$^c$ & 0.63 & $\times$ & -- " --\\
  Dec~13 04:30 & 1 & SOT-SP$^a$ & ME$^c$ & 0.63 & $\times$ & -- " --\\
 \hline
 \multicolumn{7}{|c|}{\bf FV helicity computations}\\
  Dec~11 17:00 & 1 & \CFITsc\ $\vB$ & $\times$ & 0.66 & magenta outline & \href{ss:data_fv}{Sect.~\ref{ss:data_fv}}\\
  Dec~12 20:30 & 1 & \CFITsc\ $\vB$ & $\times$ & 0.63 & $\times$ & -- " --\\
  Dec~13 04:30 & 1 & \CFITsc\ $\vB$ & $\times$ & 0.63 & $\times$ & -- " --\\
 \multicolumn{7}{|c|}{\bf \CBFF\ helicity computation}\\
  Dec~11 17:00 & 1 & \CFITsc\ $\vB$ at $z=0$ & $\times$ & 0.66 & magenta outline & \href{ss:data_fv}{Sects.~\ref{ss:data_fv}} \& \ref{ss:data_cb16}\\
  Dec~12 20:30 & 1 & \CFITsc\ $\vB$ at $z=0$ & $\times$ & 0.63 & $\times$ & -- " --\\
  Dec~13 04:30 & 1 & \CFITsc\ $\vB$ at $z=0$ & $\times$ & 0.63 & $\times$ & -- " --\\
 \multicolumn{7}{|c|}{\bf \CBSP\ helcity computation}\\
  Dec~11 03:10 -- Dec~13 16:21 & 16 & SOT-SP $\Bz$ & NPFC$^b$ & 0.31 & yellow outline$^e$ & \href{ss:data_cb16}{Sect.~\ref{ss:data_cb16}}\\
 \multicolumn{7}{|c|}{\bf FI helcity flux computation}\\
  Dec~11 12:09 -- Dec~13 12:59 & 1150 & SOT-NFI$^d$ $\Blos$ & $\times$ & 0.15 & green outline & \href{ss:data_fi}{Sect.~\ref{ss:data_fi}}\\
\lasthline
\label{tab:data}
\end{tabular}
\footnotesize{
~\\ \vspace{-10pt} \flushleft
$^a$ \url{https://csac.hao.ucar.edu/sp_data.php}\\
$^b$ Non-Potential magnetic Field Calculation (NPFC) method \citep{2005ApJ...629L..69G,2006SoPh..237..267M}\\
$^c$ Minimum-energy (ME) method \citep[]{1994SoPh..155..235M,2006SoPh..237..267M}\\
$^d$ \cite{2008PFR.....2S1009I}\\
$^e$ For December 11 data only\\
}
\end{table}

\subsection{Effect of NFI data calibration on FI computations}
\label{ss:calib_append}

In order to compute the photospheric helicity flux, three main steps are to be undertaken: calibration of $\Blos$ (or not), inversion of flux transport velocities, and helicity flux computation. From the results presented in \href{ss:discussion_fi}{Sect.~\ref{ss:discussion_fi}}, we know that the helicity flux computation itself (step three above) is not a source of major discrepancies, as the retrieved (accumulated) fluxes with the \FEP\ and \FYL\ methods are fairly consistent.

\begin{figure}[ht]
  \centering
  \includegraphics[width=0.9\textwidth]{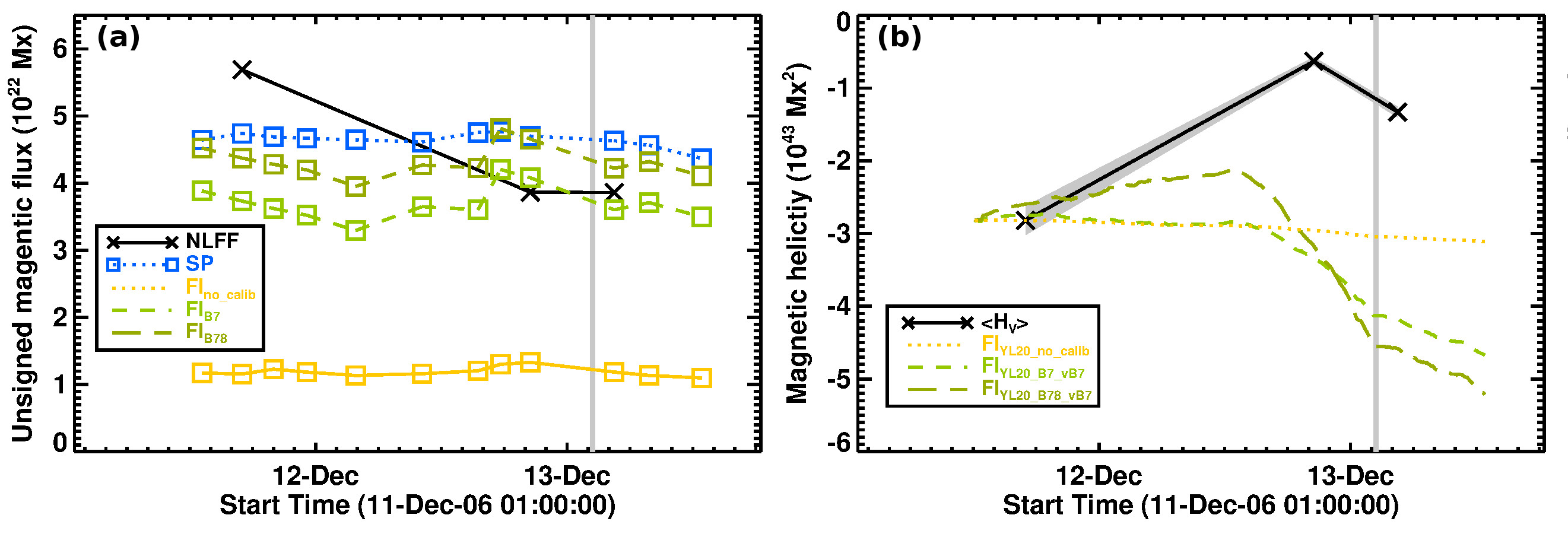}
  \caption{Magnetic and (accumulated) helicity fluxes based on different calibration products. (a) Unsigned magnetic fluxes. Estimates based on calibrated NFI $\Blos$ excluding (including) sunspot areas are shown as light green dashed (dark green long-dashed) lines, and labeled ``B7'' ("B78"). For comparison, fluxes based on the non-calibrated data are shown in yellow (labeled "no\_calib"), based on SP $\Bz$ in blue, and based on \CFITsc\ $\Bz$ in black. (b) theoretical curve for $\Hacc$ when using the mean estimate of $\Hdefm$ on December~11 as a reference level, based on \FYL\ computations (labeled ``FI$_{{\rm YL}20}$''), using DAVE velocities deduced from B7-data. For comparison, $\Hdefm$ is shown as black crosses (solid line). The vertical gray-shaded band indicates the impulsive phase of the X3.4 flare.}
  \label{fig:nfi_calib}
\end{figure}

Using the SP data as a reference \cite{2007PASJ...59S.619C} suggest to calibrate NFI data (step one mentioned above) in the following way. In regions outside the sunspots and penumbral regions, a linear relation between the circular polarization and $\Blos$ is used (their Eq.~7). The resulting calibrated $\Blos$ (light green dashed line labeled "B7" in \href{fig:nfi_calib}{Fig.~\ref{fig:nfi_calib}(a)}) yields total unsigned fluxes on the order of $\approx80\%$, on average, compared to the SP flux (blue dotted line), and about 66\%, 106\%, and 93\% of the \CFITsc\ lower boundary fluxes (black crosses) on December~11, 12 and 13, respectively. Additionally, in umbral regions a first-order polynomial can be adopted in order to model the reversal of the polarization signal over field strength (their Eq.~8). The resulting calibrated $\Blos$ (dark green long-dash, labeled "B78") yields a total unsigned flux of $\gtrsim85\%$, on average, in comparison to that of SP, and about 77\%, 120\% and 109\% of the \CFITsc\ lower boundary flux for the December 11, 12, and 13 snapshot, respectively. In comparison, the non-calibrated NFI data (yellow solid line) recovers only about 25\% of the SP unsigned magnetic flux, on average, i.e., about 20\%, 34\% and 31\% of the \CFITsc\ lower boundary flux for the December 11, 12, and 13 snapshot.

Using the non-calibrated $\Blos$ as the basis for DAVE-velocity and subsequent helicity flux computations, one obtains a predominantly negative rate of magnetic helicity injection through the photosphere during most of the considered time interval ($\propto10^{36}$\,\mxmxs; not shown explicitly) and the resulting accumulation of coronal helicity during the considered time period of $\Hacc=-2.9\times10^{42}$~\mxmx\ (see yellow dotted curve in \href{fig:nfi_calib}{Fig.~\ref{fig:nfi_calib}(b)}). Since this result, based on the non-calibrated $\Blos$, is in overall agreement with the corresponding estimates from earlier studies \citep[e.g.,][]{2008ApJ...682L.133Z,2010ApJ...720.1102P}, we must therefore question the validity of those, as well as their interpretation. This includes, e.g., earlier conclusions on $\Hacc$ to represent only a minor contribution to the coronal helicity budget (being about an order of magnitude larger in amplitude).

In order to investigate the effect of step one (calibration of $\Blos$) to the retrieved helicity flux, we compute it based on the two aforementioned data sets (B7 and B78), using $\vu$ deduced using DAVE from B7. For completeness, we note here that we do not deduce DAVE flux transport velocities from the B78 data, as discontinuities spatially associated with saturation effects, yield unrealistic velocity estimates. Though not explicitly shown, we note here that the relative helicity fluxes computed from the calibrated NFI data (both, B7 and B78, with DAVE applied to the B7 data) are remarkably higher (by a factor of $\gtrsim10$) than that based on the non-calibrated data, as is the total accumulated helicity. More precisely, we estimate a total of $\Hacc=-18.5\times10^{42}$~\mxmx\ and $\Hacc=-23.9\times10^{42}$~\mxmx\ for the considered time period, from the  B7 and B78 data sets, respectively (cf.~green dashed and long-dashed curves in \href{fig:nfi_calib}{Fig.~\ref{fig:nfi_calib}(b)}), respectively). Furthermore, instead of a rather flat time profile as for the non-calibrated data (yellow dotted curve in \href{fig:nfi_calib}{Fig.~\ref{fig:nfi_calib}(b)}), we find a time evolution drastically different, namely $\Hacc$ decreasing until about noon on December~12 and increasing until after the X-class flare.

\subsection{Effect of data disclosure on FI computations}
\label{ss:disclosure_append}

\begin{figure}[t]
  \centering
  \includegraphics[width=0.9\textwidth]{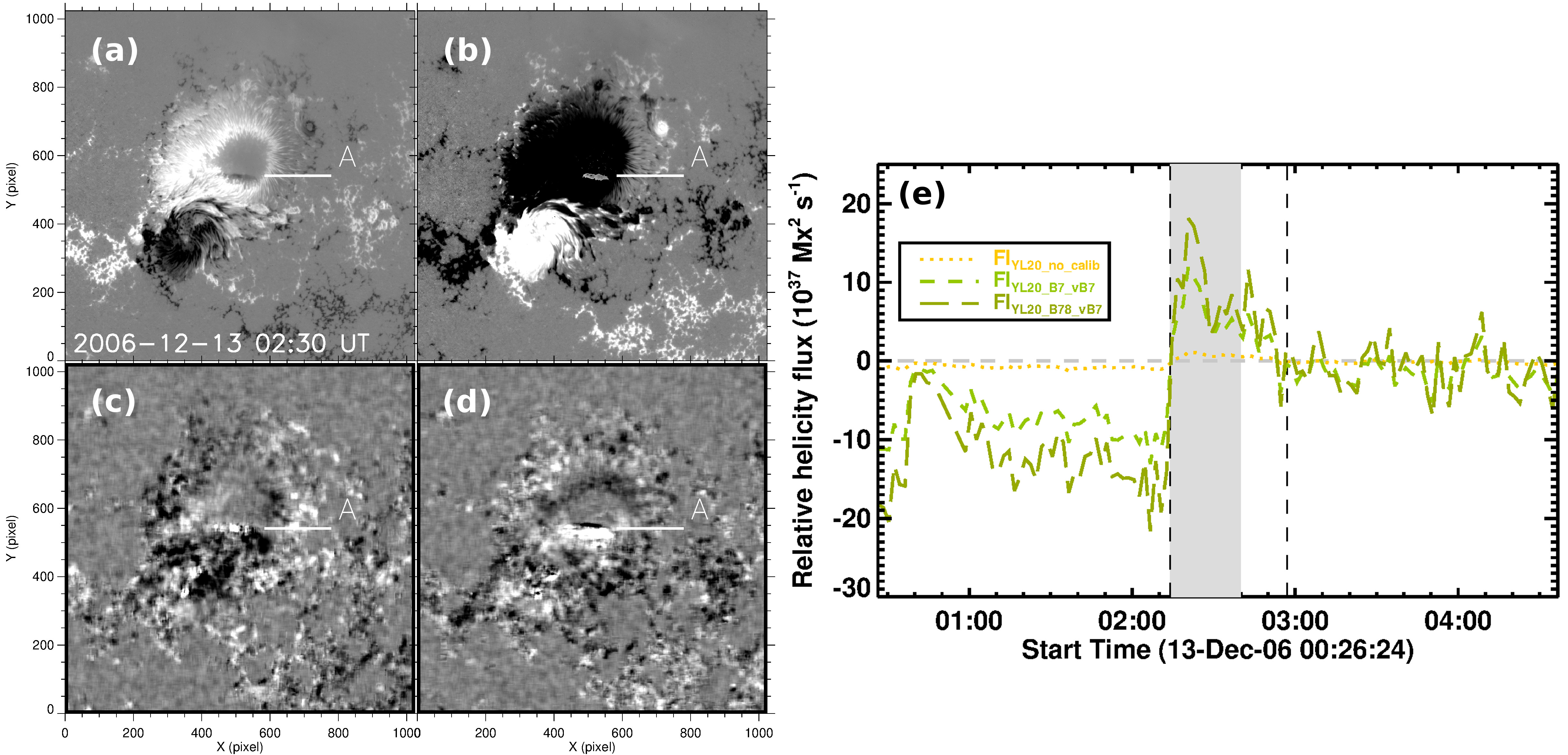}
  \caption{Magnetic field data and flux transport velocity during the flare. Calibrated NFI (a) Stokes-V and (b) $\Blos$ (B78) data. DAVE flux transport velocities, (c) $v_x$ and (d) $v_y$, saturated at $\pm0.2$\,km/s. The horizontal line labeled "A" in (a)--(d) points at the artificial patterns observed in the umbral area. (e) Helicity flux computations around the time of the X-class flare. Estimates based on calibrated NFI $\Blos$ excluding (including) sunspot areas are shown as light green dashed (dark green long-dashed) lines, and labeled ``B7'' ("B78"). Helicity fluxes based on the non-calibrated data are represented by the yellow dotted line (labeled "no\_calib"). The vertical dashed lines mark the time interval which was disclosed from analysis (the nominal GOES flare duration). The vertical gray-shaded band indicates the nominal impulsive phase of the flare.}
  \label{fig:nfi_disclose}
\end{figure}

Careful inspection of the (non-)calibrated NFI data shows artifacts largely co-spatial with the saturated umbra, especially during the flare (see \href{fig:nfi_disclose}{Fig.~\ref{fig:nfi_disclose}(a)--(d)}). In particular, nonphysical DAVE velocities are retrieved from the calibrated field, that are not representative of the magnitude or direction typically found in sunspots  (\href{fig:nfi_disclose}{Fig.~\ref{fig:nfi_disclose}(c)--(d)}). Therefore, we excluded the time range 13~December 02:14~UT -- 02:57~UT (the nominal GOES flare duration) from helicity flux analysis (indicated by the dashed vertical lines in  \href{fig:nfi_disclose}{Fig.~\ref{fig:nfi_disclose}(e)}, that shows a transition from negative to positive fluxes (irrespective of the particular calibration applied).

The effect on the deduced values for the total accumulated helicity, based on the B78 data set (favored due to the reasons outlined above), is marginal. We estimate $\Hacc=-23.7\times10^{42}$~\mxmx\ when excluding the nominal GOES flare duration (indicated by the vertical dashed lines in \href{fig:nfi_disclose}{Fig.~\ref{fig:nfi_disclose}(e)}), in comparison to $\Hacc=-22.6\times10^{42}$~\mxmx\ without data disclosure. Obviously, the latter estimate is slightly smaller due to the transition of the sudden transition of the helicity flux to positive values during the nominal GOES flare duration.

\section{NLFF modeling}
\label{s:nlff_append}
\setcounter{table}{0}
\renewcommand{\thetable}{B\arabic{table}}
\setcounter{equation}{0}
\renewcommand{\theequation}{B\arabic{equation}}

The NLFF code CFIT \citep{2007SoPh..245..251W} uses the Grad-Rubin method to solve for the coronal magnetic field, $\vB$, and the force-free parameter, $\aff$, where $\vB$ and $\aff$ are related through
\begin{linenomath*}
\begin{eqnarray}
	\curlB=\aff\vB \text{\hspace{0.5cm}and\hspace{0.5cm}} \vB\cdot\nabla\aff=0,
	\label{eq:ff}
\end{eqnarray}
\end{linenomath*}
As an input the method requires the vertical magnetic field component, $B_z$ over both magnetic polarities, and the specification of $\aff$ over one magnetic polarity. This is equivalent to a corresponding specification of the normal component of the electric current density, $J_z$, since $\vJ=\aff\vB/\mu_0$.

CFIT solves for $\vB$ and $\vJ$ iteratively, so that at the $i$th iteration
\begin{linenomath*}
\begin{equation}
	\vB^{i-1} \cdot \nabla \aff^i = 0 \mathrm{~~~and~~~} \nabla \times \vB^i = \aff^i\, \vB^{i-1}
\label{eq:bi}
\end{equation}
\end{linenomath*}
is solved \citep{2007SoPh..245..251W}. Here, given $\vB^{i-1}$, $\vB^i$ is solved for, subject to $\divB^i=0$, as well as the boundary conditions $\aff^i$ and $\Bz^i$ at $z=0$. The start equilibrium, $\vB^0$, is specified by a potential field matching the boundary conditions on $\Bz$, hence implying $\vJ^0=0$.

Grad-Rubin methods, however, have been found to deliver inconsistent results for NLFF reconstructions if relying on the boundary conditions on $\aff$ specified on one magnetic polarity only; either positive or negative \citep[][]{2008ApJ...675.1637S,2009ApJ...696.1780D}. In other words, two very different NLFF solutions may be obtained, given a single vector magnetogram. In order to resolve this inherent inconsistency, \cite{2009ApJ...700L..88W} expanded the original CFIT method by a ``self-consistency'' procedure (``\CFITsc'', hereafter).

The \CFITsc\ method is favored in the present work because of its advantageous properties, including its strict convergence when applied to solar data, hence providing a single (average) solution to the force-free problem, as explained in the following \citep[for more details see][]{2009ApJ...700L..88W,2011ApJ...728..112W}.

First, CFIT solutions are constructed for the boundary conditions on $\aff$ specified once from the positive (P solution) and once from the negative (N solution) magnetic polarity field. To reduce the effects of nonphysical currents, values of $\aff$ are used only if the signal-to-noise ratio exceeds a certain threshold and are set to zero otherwise (the so-called ``censoring''). The computational volume is defined as a uniform three-dimensional Cartesian grid (i.e., solar curvature is ignored) with equal grid spacing in each dimension. The update of $\vB$ via \href{eq:bi}{Eq.~(\ref{eq:bi})} is achieved by solving the Poisson equation for the corresponding vector potential using a two-dimensional Fourier Transform method, so that the solution is correspondingly periodic in $x$ and $y$. A mapping of $\aff$ along of model field lines in the P solution then allows to define updated boundary values in the negative polarity domain, and vice versa. Consequently, the two solutions define a new set of boundary conditions on $\aff$ over the entire vector magnetogram. Second, based on probability theory, the most probable value of $\aff$ is determined, by averaging the values of $\aff$, weighted by the uncertainties, from the original boundary values and those resulting from the mappings of the P and N solutions \citep{2009ApJ...700L..88W}. Finally, steps one and two above are iterated (self-consistency cycling), in order to obtain a self-consistent solution 
\citep[for a further recent work see][]{2020SoPh..295...97M}. 

\subsection{Quality of the \CFITsc\ models}
\label{s:quality_append}

In order to quantify the force-freeness of the obtained \CFITsc\ solutions, we use the current-density-weighted (sine of the) angle between the modeled magnetic field and electric current density, ($\rm CWsin$) $\tj$, \citep[][]{2006SoPh..235..161S} and find values ($\lesssim0.4$) $\lesssim25^\circ$ for all \CFITsc\ solutions (see last two columns in \href{tab:metrics}{Table~\ref{tab:metrics}}).

In order to assess the quality of the solenoidal condition in the \CFITsc\ solutions we use a measure for local derivations of solenoidality within the model volume, in the form of the volume-averaged fractional flux ($\avfi{(\vB)}$; see \cite{2000ApJ...540.1150W} and \cite{2020arXiv200808863G} for a most recent dedicated analysis). For all \CFITsc\ models, we find $\avfi{(\vB)}\lesssim2.0\times10^{-4}$ (see third-last column \href{tab:metrics}{Table~\ref{tab:metrics}}).

The applicability of FV methods depends critically on how well $\divB=0$ and $\divBp=0$ are fulfilled. A corresponding quality measure, alternative to $\avfi{(\vB)}$ discussed above, has been introduced by \cite{2013A&A...553A..38V}, in the form of the ratio $\Ediv/\E$, where
\begin{linenomath*}
\begin{eqnarray}
\E &=&\frac{1}{2\mu_0}\intv B^2 \dv = \Ep + \Ej \nonumber \\
  &=&\Eps+\Ejs+\Epns+\Ejns+\Emix. 
\label{eq:e_i}
\end{eqnarray}
\end{linenomath*}

\begin{table}[ht]
\caption{Quality metrics of the \CFITsc\ NLFF fields. Columns indicate the potential solenoidal ($\Eps$), current-carrying solenoidal ($\Ejs$), potential nonsolenoidal ($\EdivBp$), current-carrying nonsolenoidal ($\EdivBJ$), and mixed nonsolenoidal ($\Emix$) energy contributions normalized to the corresponding total energy $\E$ (see respective column in \href{tab:nlff_modeling}{Table~\ref{tab:nlff_modeling}}). Last columns include the average fractional flux ($\avfi{(\vB)}$), $\tj$, and the current-weighted value of $\sin\tj$, CWsin, where $\tj$ is the average angle between $\vB$ and the electric current density, $\vJ$. For a perfectly force-free solution, $\tj=0^\circ$.}
\centering
\footnotesize
\strtable
\begin{tabular}{@{~}c  | c@{\quad} c@{\quad} c@{\quad} c@{\quad} c@{\quad} | c@{\quad} c@{\quad} c@{\quad} }
\hline
   Date \& Time  & $\Eps/\E$ &$\Ejs/\E$ &$\EdivBp/\E$ & $\EdivBJ/\E$& $\Emix/\E$ & $\avfi{(\vB)} \times 10^{5}$ & CWsin & $\tj [^\circ]$\\
 \hline
  11 Dec 17:00~UT &  0.88 &  0.12 & 2.47e-04 & 1.82e-04 & 2.17e-03 & 18.22 & 0.414 & 24.460 \\
  12 Dec 20:30~UT &  0.98 &  0.02 & 1.01e-04 & 7.88e-05 & 3.24e-05 & ~8.93 & 0.380 & 22.354 \\
  13 Dec 04:30~UT &  0.94 &  0.06 & 8.20e-05 & 9.56e-05 & 1.40e-03 & ~8.82 & 0.341 & 19.950 \\
\lasthline
\label{tab:metrics}
\end{tabular}
\end{table}

Here, $\Ep$ and $\Ej$ are the energies of the potential and current-carrying magnetic field, respectively. $\Eps$ and $\Ejs$ are those of the potential and current-carrying solenoidal magnetic field components, respectively, where $\Ejs$ is the free energy in an ideal (i.e., fully solenoidal) solution. Furthermore, $\Epns$ and $\Ejns$ are the (spurious) energies of the corresponding non-solenoidal components, and $\Emix$ corresponds to all cross terms \citep[see Eq.~(8) in][for the detailed expressions]{2013A&A...553A..38V}. Except for $\Emix$, all contributions to \href{eq:e_i}{Eq.~(\ref{eq:e_i})} are positive definite. For a perfectly solenoidal field, one finds $\Eps=\Ep$, $\Ejs=\Ej$, and $\Epns=\Ejns=\Emix=0$. The sum of the non-solenoidal contributions to the total energy is $\Ediv/E$, where $\Ediv=(\Epns+\Ejns+|\Emix|)$.

In the proof-of-concept study by \cite{2016SSRv..201..147V}, based on solar-like numerical experiments, it was suggested that only for input fields achieving $\Edivprime\lesssim0.1$ a reliable helicity computation may be expected. Dedicated follow-up studies suggested an even lower threshold for solar applications \citep[$\Edivprime\lesssim0.05$;][]{2019ApJ...880L...6T}, necessarily related to minimal non-solenoidal contributions to the free magnetic energy \citep[$\Emixprime\lesssim0.4$;][]{2020A&A...643A.153T}. In the present work, all considered NLFF fields have a solenoidal level well below these thresholds, with $\Ediv/E<0.01$ and $\Emixprime<0.1$ (cf.\ numbers listed in \href{tab:metrics}{Table~\ref{tab:metrics}}). Thus, we may assume a correspondingly small error in the helicity computations.

Furthermore, based on the energy decomposition above, we find that the \CFITsc\ model magnetic fields on December 12 and 13 are found closer to a potential field state than that on December 11, as revealed by the relatively higher values of $\Eps/\Etot$ in \href{tab:metrics}{Table~\ref{tab:metrics}}. 

\section{(Accuracy of) FV and CB helicities}
\setcounter{table}{0}
\renewcommand{\thetable}{C\arabic{table}}

We list the relative helicities computed by the different FV methods (based on the \CFITsc\ solutions described in \href{ss:data_fv}{Sect.~\ref{ss:data_fv}}) in \href{tab:fv_methods_detailed}{Table~\ref{tab:fv_methods_detailed}}. The accuracy with which the vector potentials $\vA$ and $\vAp$ reproduce the respective input magnetic field, $\vB$ and $\vBp$, are listed for each of the considered FV methods in \href{tab:ei_hi}{Table~\ref{tab:ei_hi}}. The physical quantities deduced from the \CBSP\ computations, i.e. the CB method applied to the 16 available SOT-SP vector magnetograms described in \href{ss:data_cb16}{Sect.~\ref{ss:data_cb16}}, are displayed in \href{tab:cbsp_detailed}{Table~\ref{tab:cbsp_detailed}}.

\begin{table}[h]
\caption{FV computations. Relative helicities $\Hdef$, $\HdefJ$, and $\HdefJP$ obtained from four individual FV helicity computation methods (see \href{ss:fv_methods}{Sect.~\ref{ss:fv_methods}}) are given in units of $10^{43}$\,\mxmx.}
 \centering
 \footnotesize
 \strtable
 \begin{tabular}{
 @{~}c | @{~}c |
 c@{\quad} c@{\quad} c@{\quad} |
 c@{\quad} c@{\quad} 
    }
\firsthline
Method (as in  & Date \& Time & \multicolumn{3}{c|}{Relative helicities} & \multicolumn{2}{c}{Intensive quantities} \\
     \cline{3-7}
\href{ss:fv_methods}{Sect.~\ref{ss:fv_methods}} and \href{tab:methods}{Table~\ref{tab:methods}}) & ~ & $\Hdef$ & $\HdefJ$ & $\HdefJP$ & $\THdef$ & $\hjprime$ \\
 \firsthline
 \sJT{} & 11 Dec 17:00~UT & -3.1472 & -0.2626 & -1.8847 & -0.0389 & 0.0834\\
 \sJT{} & 12 Dec 20:30~UT & -0.6780 & -0.0040 & -0.6741 & -0.0182 & 0.0058\\
 \sJT{} & 13 Dec 04:30~UT & -1.4449 & -0.0405 & -1.4043 & -0.0388 & 0.0281\\
 \sSY{} & 11 Dec 17:00~UT & -2.6896 & -0.2634 & -2.4263 & -0.0332 & 0.0979\\
 \sSY{} & 12 Dec 20:30~UT & -0.5832 & -0.0049 & -0.5783 & -0.0156 & 0.0083\\
 \sSY{} & 13 Dec 04:30~UT & -1.2693 & -0.0417 & -1.2276 & -0.0341 & 0.0328\\
 \sKM{} &11 Dec 17:00~UT & -2.6993 & -0.2633 & -2.4360 & -0.0334 & 0.0975\\
 \sKM{} & 12 Dec 20:30~UT & -0.6116 & -0.0051 & -0.6065 & -0.0164 & 0.0084\\
 \sKM{} & 13 Dec 04:30~UT & -1.3354 & -0.0437 & -1.2917 & -0.0359 & 0.0327\\
 \sGV{} & 11 Dec 17:00~UT & -2.6912 & -0.2629 & -2.4283 & -0.0333 & 0.0977\\
 \sGV{} & 12 Dec 20:30~UT & -0.5817 & -0.0046 & -0.5771 & -0.0156 & 0.0079\\
 \sGV{} & 13 Dec 04:30~UT & -1.2701 & -0.0416 & -1.2285 & -0.0341 & 0.0327\\
 \sSA{} & 11 Dec 17:00~UT & -2.6993 & -0.2635 & -2.4358 & -0.0334 & 0.0976\\
 \sSA{} & 12 Dec 20:30~UT & -0.5847 & -0.0049 & -0.5798 & -0.0156 & 0.0084\\
 \sSA{} & 13 Dec 04:30~UT & -1.2751 & -0.0417 & -1.2334 & -0.0342 & 0.0327\\
 \lasthline
 \label{tab:fv_methods_detailed}
 \end{tabular}
\end{table}

\begin{table}
\caption{Metrics of accuracy of vector potentials in reproducing the corresponding magnetic field. $C_{\rm Vec}$ is the vector correlation, analogous to the standard correlation coefficient for scalar functions. If two vector fields are identical, then $C_{\rm Vec}=1$. $C_{\rm CS}$ is a measure of the angular differences of the vector fields. For parallel fields, $C_{\rm CS}=1$, and is $-1$ for anti-parallel fields. $\epsN$ and $\epsM$ are the complements of the normalized and mean vector error, respectively. $\epsilon_{\rm E}$ gives the energy ratio, i.e., the ratio of the total energy with respect to the input field. For the computation of these metrics the standard numerical prescriptions as described in Appendix~A of \cite{2013A&A...553A..38V} were applied.}
 \centering
 \footnotesize
 \strtable
 \begin{tabular}{
 @{~}c | @{~}c |
 c@{\quad} c@{\quad} c@{\quad} c@{\quad} c@{\quad} |
 c@{\quad} c@{\quad} c@{\quad} c@{\quad} c@{\quad} }
\firsthline
Method (as in  & Date \& Time & \multicolumn{5}{c|}{$\vBp$ vs $\Nabla \times\vAp$} &
     \multicolumn{5}{c}{$\vB$ vs $\Nabla \times\vA$} \\
     \cline{3-12}
\href{ss:fv_methods}{Sect.~\ref{ss:fv_methods}} and \href{tab:methods}{Table~\ref{tab:methods}}) & & $C_{\rm Vec}$ &$C_{\rm CS}$ &$\epsN$ & $\epsM$& $\epsilon_{\rm E}$ & $C_{\rm Vec}$ &$C_{\rm CS}$ &$\epsN$ & $\epsM$& $\epsilon_{\rm E}$ \\
 \firsthline
 \sJT{} & 11 Dec 17:00~UT & 0.6685 & 0.6283 & 0.0058 & -0.6641 & 2.2166 & 0.6909 & 0.6473 & 0.0693 & -0.6612 & 2.0819 \\
 \sJT{} & 12 Dec 20:30~UT & 0.6429 & 0.6163 & -0.1707 & -0.7033 & 2.4263 & 0.6461 & 0.6182 & -0.1667 & -0.7059 & 2.4118 \\
 \sJT{} & 13 Dec 04:30~UT & 0.6339 & 0.6150 & -0.1860 & -0.7055 & 2.4841 & 0.6451 & 0.6251 & -0.1604 & -0.7105 & 2.4053 \\
 \sSY{} & 11 Dec 17:00~UT & 0.9996 & 0.9990 & 0.9758 & 0.9604 & 1.0011 & 0.9995 & 0.9987 & 0.9738 & 0.9552 & 0.9985\\
 \sSY{} & 12 Dec 20:30~UT & 0.9995 & 0.9992 & 0.9724 & 0.9594 & 1.0011 & 0.9995 & 0.9988 & 0.9681 & 0.9523 & 1.0005\\
 \sSY{} & 13 Dec 04:30~UT & 0.9995 & 0.9992 & 0.9722 & 0.9595 & 1.0011 & 0.9995 & 0.9988 & 0.9686 & 0.9525 & 0.9992 \\
 \sKM{} & 11 Dec 17:00~UT & 0.9999 & 1.0000 & 0.9985 & 0.9995 & 0.9994 & 0.9998 & 1.0000 & 0.9981 & 0.9994 & 0.9995 \\
 \sKM{} & 12 Dec 20:30~UT & 0.9999 & 1.0000 & 0.9536 & 0.9533 & 1.0885 & 0.9999 & 1.0000 & 0.9535 & 0.9533 & 1.0887 \\
 \sKM{} & 13 Dec 04:30~UT & 0.9999 & 1.0000 & 0.9536 & 0.9533 & 1.0887 & 0.9998 & 1.0000 & 0.9534 & 0.9533 & 1.0889 \\
 \sGV{} & 11 Dec 17:00~UT & 0.9999 & 1.0000 & 0.9981 & 0.9990 & 0.9980 & 1.0000 & 1.0000 & 0.9987 & 0.9995 & 0.9987 \\
 \sGV{} & 12 Dec 20:30~UT & 1.0000 & 1.0000 & 0.9986 & 0.9990 & 0.9985 & 1.0000 & 1.0000 & 0.9991 & 0.9997 & 0.9990 \\
 \sGV{} & 13 Dec 04:30~UT & 1.0000 & 1.0000 & 0.9987 & 0.9991 & 0.9985 & 1.0000 & 1.0000 & 0.9992 & 0.9997 & 0.9990 \\ 
 \sSA{} & 11 Dec 17:00~UT & 0.9999 & 1.0000 & 0.9988 & 0.9996 & 1.0004 & 0.9999 & 1.0000 & 0.9982 & 0.9994 & 0.9995 \\
 \sSA{} & 12 Dec 20:30~UT & 1.0000 & 1.0000 & 0.9993 & 0.9998 & 1.0002 & 1.0000 & 1.0000 & 0.9989 & 0.9996 & 0.9998 \\
 \sSA{} & 13 Dec 04:30~UT & 1.0000 & 1.0000 & 0.9993 & 0.9998 & 1.0002 & 0.9999 & 1.0000 & 0.9989 & 0.9996 & 0.9995 \\
 \lasthline
 \label{tab:ei_hi}
 \end{tabular}
\end{table}

\begin{table}
\caption{
\CBSP\ results. Unsigned magnetic flux ($\phi$), total connected flux ($\phi_c$), relative helicity ($\HdefCB$) and corresponding error ($\Delta\HdefCBlh$), right-handed helicity ($\HdefCBrh$) and corresponding error ($\Delta\HdefCBrh$), left-handed helicity ($\HdefCBlh$) and corresponding error ($\Delta\HdefCBlh$), total ($\E$) and potential ($\Ep$) magnetic energy, along with error $\Delta E$ in $E$. Magnetic fluxes are given in units of $10^{22}$\,\mx, energies in units of $10^{33}$\,\erg, and helicities in units of $10^{43}$\,\mxmx.}
 \centering
 \footnotesize
 \strtable
 \begin{tabular}{
 @{~}c |
 c@{\quad} c@{\quad} |
 c@{\quad} c@{\quad} c@{\quad} c@{\quad} c@{\quad} c@{\quad} |
 c@{\quad} c@{\quad} c@{\quad}
    }
\firsthline
Date \& Time & \multicolumn{2}{c|}{Magnetic fluxes} & \multicolumn{6}{c|}{Relative helicities} & \multicolumn{3}{c}{Magnetic energies} \\
     \cline{2-12}
  ~ & $\phi$ & $\phi_c$ & 
  $\HdefCB$ & $\Delta\HdefCB$ & 
  $\HdefCBrh$ & $\Delta\HdefCBrh$ & $\HdefCBlh$ & $\Delta\HdefCBlh$ &
  $E$ & $ \Delta E$ & $\Ep$\\
  ~ & & & 
    & ($\times10^{-2}$) & 
    & ($\times10^{-2}$) & & ($\times10^{-2}$) &
    & ($\times10^{-3}$) & \\
 \firsthline
	11~Dec 03:10:04 & 4.8571 & 3.460 & -1.6366 & 1.8668 & 0.2794 & 1.0956 & -1.9159 & 1.5115 & 2.2773 & 4.3920 & 1.8455 \\
	11~Dec 08:00:04 & 4.9060 & 3.412 & -0.7902 & 4.1563 & 0.8691 & 2.5952 & -1.6593 & 3.2465 & 2.2166 & 5.1245 & 1.8415 \\
	11~Dec 11:10:06 & 4.8468 & 3.416 & -1.9559 & 1.6046 & 0.1362 & 0.5837 & -2.0921 & 1.4946 & 2.3332 & 4.5147 & 1.8410 \\
	11~Dec 13:10:09 & 4.6428 & 3.148 & -1.5726 & 2.0504 & 0.2121 & 1.0199 & -1.7847 & 1.7788 & 2.1877 & 4.3450 & 1.8031 \\
	11~Dec 17:00:08 & 4.7401 & 3.228 & -1.6789 & 2.8096 & 0.2889 & 1.3113 & -1.9679 & 2.4848 & 2.2087 & 5.4217 & 1.8251 \\
	11~Dec 20:00:05 & 4.6891 & 3.220 & -0.1214 & 4.6167 & 1.2331 & 3.0320 & -1.3546 & 3.4815 & 2.1001 & 6.1313 & 1.7914 \\
	11~Dec 23:10:05 & 4.6676 & 3.212 & -1.7192 & 1.7963 & 0.1828 & 0.7172 & -1.9020 & 1.6470 & 2.2220 & 4.3465 & 1.7884 \\
	12~Dec 03:50:05 & 4.6455 & 3.116 & -1.7519 & 1.7362 & 0.1495 & 0.6102 & -1.9015 & 1.6255 & 2.2010 & 4.2237 & 1.7664 \\
	12~Dec 10:10:08 & 4.6140 & 3.060 & -1.1519 & 2.5019 & 0.3321 & 1.0915 & -1.4840 & 2.2513 & 2.0862 & 4.9743 & 1.7227 \\
	12~Dec 15:30:08 & 4.7559 & 3.168 & -0.9320 & 3.8538 & 0.4975 & 2.2275 & -1.4295 & 3.1448 & 2.0300 & 4.9267 & 1.7116 \\
	12~Dec 17:40:05 & 4.7669 & 3.216 & -1.1977 & 2.3044 & 0.3575 & 1.3592 & -1.5552 & 1.8609 & 2.0813 & 4.4163 & 1.7034 \\
	12~Dec 20:30:05 & 4.7015 & 3.148 & -0.3878 & 3.6809 & 0.7175 & 2.8076 & -1.1053 & 2.3803 & 1.9293 & 4.6931 & 1.6680 \\
	13~Dec 04:30:05 & 4.6319 & 3.016 & -1.1493 & 1.1832 & 0.1131 & 0.5406 & -1.2624 & 1.0525 & 1.9230 & 3.9054 & 1.6160 \\
	13~Dec 07:50:05 & 4.5646 & 2.976 & -1.1658 & 1.0349 & 0.0827 & 0.3717 & -1.2485 & 0.9659 & 1.8930 & 4.0902 & 1.5910 \\
	13~Dec 12:51:04 & 4.3675 & 2.716 & -1.0129 & 1.1522 & 0.0978 & 0.3922 & -1.1107 & 1.0834 & 1.7029 & 3.3812 & 1.4607 \\
	13~Dec 16:21:04 & 4.4393 & 2.828 & -1.0250 & 1.1816 & 0.1537 & 0.6383 & -1.1786 & 0.9943 & 1.7662 & 3.9294 & 1.4953 \\
 \lasthline
 \label{tab:cbsp_detailed}
 \end{tabular}
\end{table}

\clearpage
\bibliography{bibliography} 
\end{document}